\documentclass[aps,prb,twocolumn,amsmath,amssymb]{revtex4-2}
\usepackage{graphicx} 
\usepackage{bm}
\usepackage{braket}
\usepackage{mathtools}
\usepackage{hyperref}
\hypersetup{colorlinks=true, urlcolor=blue, linkcolor=blue, citecolor=blue}

\begin{document}

\title{Superconductivity in doped spin multimer systems}

\author{Ritsuki Hirabayashi, Masataka Kakoi, Ryota Ueda, Kazuhiko Kuroki, and Tatsuya Kaneko}

\affiliation{Department of Physics, The University of Osaka, Toyonaka, Osaka 560-0043, Japan}

\begin{abstract}
Binding energy, which quantifies pair formation, is a key factor in the emergence of superconductivity. 
Here, we show that even when multiple spins are complexly coupled, hole-doped systems can be mapped, in the strong-binding-energy limit, onto an effective hardcore boson model equivalent to the XXZ model, and exhibit promising signatures of superconductivity. 
We demonstrate this theory analytically and numerically in the double Kondo lattice model. 
Using the density-matrix renormalization group method, we show that a pairing state is maintained via a crossover even for parameters away from the strong-coupling regime. 
Additionally, we find that once binding energies are sufficiently generated, pair correlations develop similarly regardless of the details of local spin correlations. 
Our findings provide useful insights into the mechanism of superconductivity. 
\end{abstract}

\maketitle

\textit{Introduction.} Elucidating and proposing mechanisms of high-temperature superconductivity (SC) is a central challenge in condensed matter physics~\cite{EDagotto1994,JOrenstein2000,PLee2006,BKeimer2015}. 
One key factor for SC is how two carriers bind to form a pair; the binding energy quantifies the strength of pair formation. 
A pairing state with a large binding energy is characterized by the Bose-Einstein condensation (BEC) of tightly bound pairs. 
Even when the binding energy is not particularly large, a SC state can still be maintained due to a crossover from the BEC regime, such as in $s$-wave SC in the attractive Hubbard model~\cite{PNozieres1985,RMicnas1990,MRanderia2014}. 
Therefore, research on systems with guaranteed binding energies can provide valuable insights for the design of SC materials. 

A pairing state across two magnetically coupled layers can also be discussed in this context. 
When electrons are strongly correlated, doped holes prefer to form interlayer pairs to avoid breaking spin-singlet bonds, which leads to a binding energy~\cite{EDagotto1992,SHirthe2023}. 
Inspired by this idea, the pairing properties in strongly correlated bilayers and ladders have been investigated using various theoretical techniques~\cite{HTsunetsugu1994,CHayward1995,MTroyer1996,GSierra1998,TSiller2001,ABohrdt2022,XZQu2024,JChen2024,HLange2024,HSchlomer2024,HSchlomer2024_2,DCLu_arXiv,YYang2026,HLange_arXiv}. 
The recent discovery of high-temperature SC in multilayer nickelates~\cite{HSun2023,NWang2024,HSakakibara2024_4310,YZhu2024,EKKo2025,GZhou2025} has heightened the importance of the pairing mechanism in the bilayer structure~\cite{KKuroki2002,TMaier2011,MDolfi2015,YShen2023,HOh2023,TKaneko2024,CLu2024_1,MKakoi2024,ZLuo2024,HYang2024,HOh2025,TKaneko2025,XZQu2025,TMaier2026,HOh2026,HOh_arXiv,HWatanabe_arXiv}. 
Pair formation in bilayer nickelates is more complicated because the nearly quarter-filled Ni $3d_{x^2-y^2}$ orbitals and nearly half-filled Ni $3d_{3z^2-r^2}$ orbitals form the electronic structure near the Fermi level~\cite{MNakata2017,ZLuo2023,QGYang2023,VChristiansson2023,YZhang2023,FLechermann2023,HSakakibara2024_327,SRyee2024,KUshio2026}. 
The interplay between two orbitals with distinct properties is likely responsible for SC in bilayer nickelates. 
Therefore, understanding pair formation and the resulting SC states in systems with multiple sources of pairing is crucial for achieving high-temperature SC in complex materials. 

In this Letter, we show that even when multiple spins are complexly coupled in multiorbital systems, correlated electrons that can be mapped onto an effective hardcore boson model underpinned by a binding energy exhibit promising signatures of SC. 
We verify this scenario in two Kondo-lattice chains coupled via an interaction between localized spins. 
We derive an integrable model describing strongly paired holes, and then numerically demonstrate that a pairing state exhibits a crossover from the BEC regime as parameters deviate from the strong-binding-energy limit, using the density-matrix renormalization group (DMRG) method. 
Moreover, we show that when two different systems have substantial binding energies that are comparable to each other, their pair correlations develop similarly regardless of the details of local spin correlations. 

\begin{figure}[b]
\begin{center}
\includegraphics[width=\columnwidth]{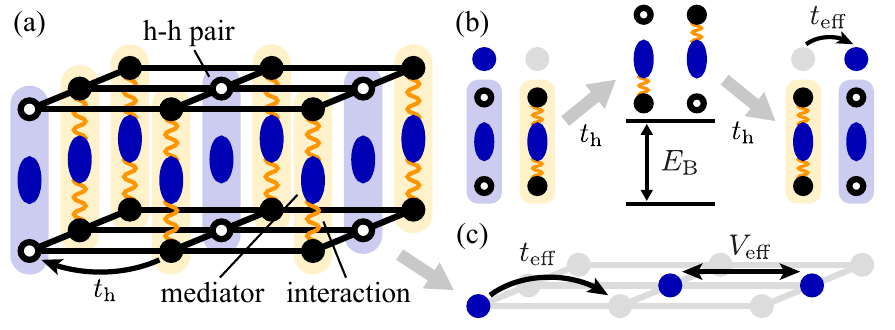} 
\caption{(a) Two layers coupled via mediators. 
(b) Second-order process that yields the effective hopping $t_{\rm eff}$ of a hole-hole (h-h) pair. 
(c) Effective hardcore boson model.} 
\label{fig1}
\end{center}
\end{figure}

\textit{Effective model based on binding energy.} First, we introduce an effective model for tightly bound pairs guaranteed by the binding energy 
\begin{equation}
E_{\rm B} = 2E_0(1) - E_0(0) - E_0(2). 
\label{eq:EB}
\end{equation}
Here, $E_0(N_h)$ is the lowest energy of the $N_h$-hole state in a unit cell. 
When $E_{\rm B}$ is much larger than the single-particle hopping $t_{\rm h}$ (i.e., $E_{\rm B} \gg t_{\rm h}$), Eq.~\eqref{eq:EB} corresponds to the change in energy during a perturbative pair-breaking process [Fig.~\ref{fig1}(b)]. 
When hole-hole pairs are represented by hardcore bosons, a system composed of tightly bound pairs (without any $N_h=1$ configuration) can be described by
\begin{equation}
\hat{H}_{\rm eff} = -t_{\rm eff} \sum_{\braket{\bm{i},\bm{j}}} \left(\hat{b}^{\dag}_{\bm{i}}\hat{b}^{}_{\bm{j}} + {\rm H.c.}\right)
+ V_{\rm eff} \sum_{\braket{\bm{i},\bm{j}}} \hat{n}^b_{\bm{i}}\hat{n}_{\bm{j}}^b
+ \epsilon_{\rm eff} \sum_{\bm{j}} \hat{n}_{\bm{j}}^b, 
\label{eq:hcb_model}
\end{equation}
where $\hat{b}^{\dag}_{\bm{j}}$ is the creation operator of a hardcore boson at site $\bm{j}$ and $\hat{n}_{\bm{j}}^b=\hat{b}^{\dag}_{\bm{j}}\hat{b}^{}_{\bm{j}}$. 
The energy of a single boson $\epsilon_{\rm eff}$ depends on system details~\cite{SM}. 
According to second-order perturbation theory, the effective hopping $t_{\rm eff}$ and the nearest-neighbor interaction $V_{\rm eff}$ are given by $V_{\rm eff} = 2t_{\rm eff} = 4 t^2_{\rm h}/E_{\rm B}$. 
For example, in the attractive Hubbard model, a strong on-site attraction $U$ yields $E_{\rm B}=U$ and $V_{\rm eff} = 2t_{\rm eff} = 4 t^2_{\rm h}/U$~\cite{PNozieres1985,RMicnas1990}. 

In the bilayer platform, the framework outlined above has been applied to the $t$-$J$ model with a strong interlayer magnetic coupling $J_{\perp}$\cite{MTroyer1996,GSierra1998,ABohrdt2021}. 
Particularly, in the mixed-dimensional (mixD) $t$-$J$ model (where the interlayer hopping $t_{\perp}$ is assumed to be zero)~\cite{ABohrdt2022,XZQu2024,JChen2024,HLange2024,HSchlomer2024,HSchlomer2024_2,HLange_arXiv,DCLu_arXiv,YYang2026}, $J_{\perp}$ becomes the pairing glue of an interlayer hole-hole pair to avoid breaking spin-singlet dimers. 
When the in-plane hopping $t_{\parallel}$ is much smaller than $J_{\perp}$, interlayer pairs guaranteed by $E_{\rm B}=J_{\perp}$ can be described by the hardcore boson model with $V_{\rm eff} = 2t_{\rm eff} = 4 t^2_{\parallel}/J_{\perp}$~\cite{MTroyer1996,ABohrdt2021,HLange2024,HSchlomer2024,HSchlomer2024_2,HLange_arXiv}. 
Even though $J_{\perp} \gg t_{\parallel}$ is not sufficiently satisfied, a crossover from a state characterized by BEC of tightly bound pairs to a spatially extended Bardeen-Cooper-Schrieffer (BCS) like state as a function of $t_{\parallel}/J_{\perp}$ has been numerically demonstrated, i.e., the emergence of SC is expected even under realistic parameters~\cite{JChen2024,HSchlomer2024,HLange_arXiv}. 

The examples mentioned above are cases where the existence of spin-singlet pairs is evident. 
However, we will demonstrate that even when spin multimers that embed spin singlets are formed from more than two spins, hole-doped systems that can be mapped onto the hardcore boson model can create signatures favorable for SC. 
For example, if two itinerant-electron layers are coupled via mediators as shown in Fig.~\ref{fig1}(a), magnetically interacting electrons and mediators form local spin multimers. 
Even in such a case, a binding-energy-based perspective on SC can be formulated. 

\begin{figure}[b]
\begin{center}
\includegraphics[width=\columnwidth]{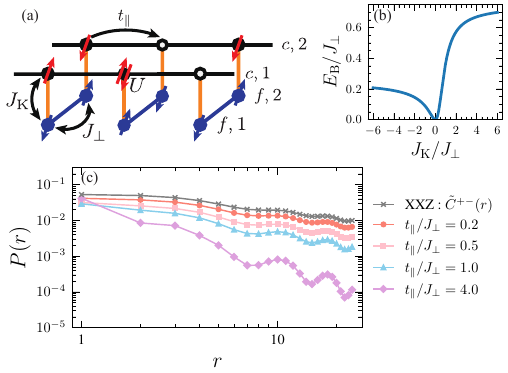} 
\caption{(a) Two-leg Kondo-Hubbard ladder and (b) binding energy $E_{\rm B}$ when $t_{\parallel}=0$.
(c) Pair correlation function $P(r)$ for $J_{\rm K}/J_{\perp}=2$ at $n_h=0.125$, where $\widetilde{C}^{+-}(r)=|\alpha|^2C^{+-}(r)$ is given by the transverse spin correlation function $C^{+-}(r)$ of the effective XXZ chain with the renormalization factor $\alpha=\cos \tfrac{\theta}{2}$.} 
\label{fig2}
\end{center}
\end{figure}

\textit{Double Kondo lattice model.} To verify the above theory in a practical model, we consider the model consisting of two Kondo-lattice layers magnetically coupled via localized spins, dubbed the double Kondo lattice model~\cite{HYang2025}. 
Here, we employ the ladder structure shown in Fig.~\ref{fig2}(a) to discuss the validity of our theory using quasi-exact numerical results obtained by the DMRG method. 
Introducing an electron-electron interaction that can realistically exist, the Hamiltonian is given by 
\begin{align}
\hat{H}
=&-t_{\parallel} \sum_{j,l,\sigma} \left( \hat{c}^{\dag}_{j,l,\sigma} \hat{c}_{j+1,l,\sigma} + {\rm H.c.} \right)
+ U \sum_{j,l} \hat{n}^c_{j,l,\uparrow} \hat{n}^c_{j,l,\downarrow}
\notag \\
&+ J_{\rm K} \sum_{j,l} \hat{\bm{S}}^c_{j,l} \cdot \hat{\bm{S}}^f_{j,l} 
+J_{\perp} \sum_{j} \hat{\bm{S}}^f_{j,1} \cdot \hat{\bm{S}}^f_{j,2}. 
\end{align}
Here, $\hat{c}_{j,l,\sigma}$ ($\hat{c}^{\dag}_{j,l,\sigma}$) annihilates (creates) an electron with spin $\sigma$~$(=\uparrow,\downarrow)$ on chain $l$~(=1,2) at site $j$, while $\hat{n}_{j,l,\sigma}^{c}=\hat{c}^{\dag}_{j,l,\sigma} \hat{c}_{j,l,\sigma}$ and $\hat{\bm{S}}_{j,l}^{c}= \sum_{\alpha,\beta}{\hat{c}_{j,l,\alpha}^{\dagger} \bm{\sigma}_{\alpha\beta} \hat{c}_{j,l,\beta}}/2$ are the number and spin operators, respectively, where $\bm{\sigma}$ is the vector of Pauli matrices. 
The operator $\hat{\bm{S}}_{j,l}^f$ describes a localized $S=1/2$ spin. 
The single-electron hopping is given by $t_{\parallel}$, and the Coulomb repulsion $U$ acts between doubly occupied electrons. 
An electron and a localized spin are coupled via $J_{\rm K}$, while localized spins are coupled via the antiferromagnetic interaction $J_{\perp}$. 
Two localized spins coupled via $J_{\perp}$ correspond to the mediator in Fig.~\ref{fig1}(a). 

When $t_{\parallel}=0$, two holes form a pair whose binding energy $E_{\rm B}$ is given by the eigenenergies in a four-site unit ($c$-$f$-$f$-$c$) cell~\cite{SM}. 
Figure~\ref{fig2}(b) shows $E_{\rm B}$ as a function of $J_{\rm K}/J_{\perp}$, which is consistent with the assessment by Yang {\it et al.} \cite{HYang2025}. 
The binding energy is generated in both the ferromagnetic ($J_{\rm K}<0$) and antiferromagnetic ($J_{\rm K}>0$) Kondo coupling regions, where $E_{\rm B} \rightarrow J_{\perp}/4$ at $J_{\rm K} \rightarrow -\infty$ while $E_{\rm B} \rightarrow 3J_{\perp}/4$ at $J_{\rm K} \rightarrow \infty$, suggesting that $J_{\rm K}>0$ is better for pair formation. 

When $E_{\rm B}$ ensures the stability of a pair, a hardcore boson in the effective system encodes whether site $j$ is in the zero-hole ($N_h=0$) or two-hole ($N_h=2$) state. 
In the $N_h=2$ state due to the absence of electrons in a unit cell, two localized spins form a spin singlet, i.e., the lowest-energy state is $\ket{g_{N_h=2}}=\ket{S^f=0}$, where $S^{\mu}$ denotes the spin quantum number in orbital $\mu$~($=c,f$). 
On the other hand, the $N_h=0$ state is configured by four spins. 
Tetramer (four-spin) states can be spanned by $\ket{S_{\rm tot}, M; S^f, S^c}$, where $S_{\rm tot}$ and $M$ denote the total spin and net magnetization in a unit cell~\cite{SM,CGC}. 
Employing this representation, the lowest-energy state for $N_h=0$ is given by 
\begin{align}
\Ket{g_{N_h=0}} &= \cos\frac{\theta}{2}\Ket{0,0;0,0} + \sin\frac{\theta}{2}\Ket{0,0;1,1}, 
\label{eq:zero_hole}
\end{align}
where $\tan\theta = \sqrt{3}J_{\rm K}/(J_{\perp}-J_{\rm K})$, see the Supplemental Material for details~\cite{SM}. 

The double Kondo lattice model can be mapped onto the hardcore boson model of Eq.~\eqref{eq:hcb_model} when $t_{\parallel}\ll E_{\rm B}$. 
The perturbation introduced by $t_{\parallel}$ yields $t_{\rm eff}$ and $V_{\rm eff}$ in Eq.~\eqref{eq:hcb_model}. 
As shown in Fig.~\ref{fig1}(b), second-order processes incorporate $N_h=1$ excited states. 
Similarly, trimer (three-spin) states can be spanned by $\ket{S_{\rm tot}, M; S^f, S^c}$. 
The lowest-energy state in the $N_h=1$ sector is 
\begin{align}
\Ket{g^{\sigma}_{N_h=1}}_{1,0} &= \cos\frac{\phi}{2}\Ket{\tfrac12,\tfrac{\sigma}{2};0,\tfrac12}_{1,0} + \sin\frac{\phi}{2}\Ket{\tfrac12,\tfrac{\sigma}{2};1,\tfrac12}_{1,0},
\label{eq:one_hole}
\end{align}
where $\sigma=\pm 1$ and $\tan\phi=\sqrt{3}J_{\rm K}/(2J_{\perp}-J_{\rm K})$~\cite{SM}. 
The subscript $(n^c_1,n^c_2)=(1,0)$ represents the position of an electron. 
Taking only the above lowest excited state into account, $t_{\rm eff}$ and $V_{\rm eff}$ for the hardcore boson model are given by 
\begin{equation}
V_{\rm eff} = 2t_{\rm eff} = \frac{4t_{\parallel}^2}{E_{\rm B}}\cos^2\frac{\phi}{2}\cos^2\left(\frac{\theta-\phi}{2}\right) . 
\end{equation}
In contrast to the mixD $t$-$J$ model, $t_{\rm eff}$ and $V_{\rm eff}$ depend on the Kondo coupling $J_{\rm K}$ included in $\theta$, $\phi$, and $E_{\rm B}$. 
The details of the derivation and more complete forms that incorporate all possible excited states are presented in the Supplemental Material~\cite{SM}. 
This mapping is applicable to any electron filling, $U\ge 0$, $J_{\perp}>0$, and $J_{\rm K}\neq0$ when $t_{\parallel} \ll E_{\rm B} (J_{\perp},J_{\rm K})$ is satisfied. 

\begin{figure*}[t]
\begin{center}
\includegraphics[width=0.9\textwidth]{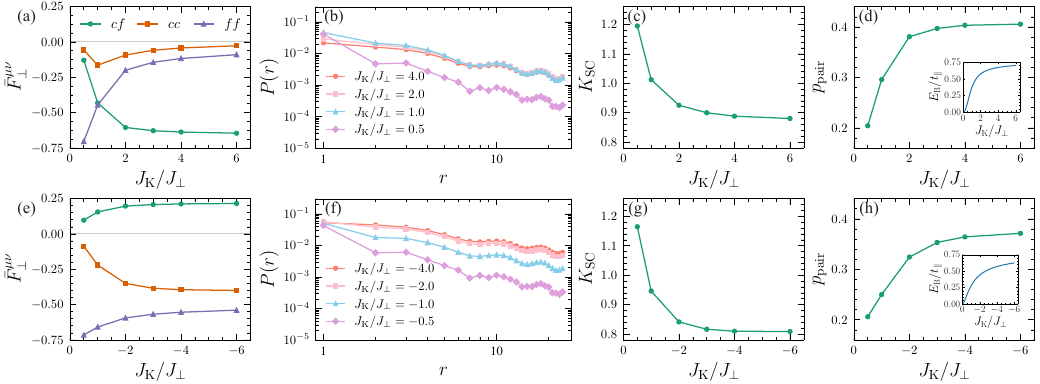} 
\caption{Comparison of the ground-state properties for $J_{\rm K}>0$ (upper panel) and $J_{\rm K}<0$ (lower panel), where $t_\parallel/J_\perp=1$ is used in (a)--(d) while $t_\parallel/J_\perp=1/3$ is used in (e)--(h) to equalize $E_{\rm B}/t_{\parallel}$ at $|J_{\rm K}|/J_{\perp} \gg 1$. 
(a), (e) Spin correlations within a unit cell $\bar{F}^{cc}_{\perp}$, $\bar{F}^{ff}_{\perp}$, and $\bar{F}^{cf}_{\perp}$, where the site-averaged values are plotted. 
(b), (f) Pair correlation function $P(r)$. 
(c), (g) Decay exponent $K_{\rm SC}$ of $P(r)$. 
(d), (h) Fraction of paired holes among all holes $p_{\rm pair}$, where the insets present the binding energies in units of $t_{\parallel}$.} 
\label{fig3}
\end{center}
\end{figure*}

\textit{Numerical demonstrations.} We numerically demonstrate the binding-energy-based pairing mechanism and the crossover from the strong-coupling ($E_{\rm B} \gg t_{\parallel}$) regime using the DMRG method~\cite{SWhite1992,SWhite1993,USchollwock2011}. 
In our calculation, the finite-size algorithm is applied to the $L_x \times 2$ site ladder with open boundary conditions. 
We present the results for $U/t_{\parallel}=8$, $L_x=48$, and $n_h=N_h/(2L_x)=0.125$, where $N_h$ is the number of holes doped into the half-filled system. 
The bond dimension is up to $m=6000$, where the maximum truncation error is less than $10^{-5}$. 
The numerical results supporting the validity of our DMRG calculations are provided in the Supplemental Material~\cite{SM}. 

We evaluate the pairing state using the pair correlation function $P(r)=\braket{ \hat{\Delta}_{j+r}^{\dagger} \hat{\Delta}_{j} }$, where $\hat{\Delta}^{\dagger}_j = ( \hat{c}^{\dagger}_{j,1,\uparrow}\hat{c}^{\dagger}_{j,2,\downarrow} - \hat{c}^{\dagger}_{j,1,\downarrow}\hat{c}^{\dagger}_{j,2,\uparrow} ) /\sqrt{2}$ is the creation operator of an interchain spin-singlet pair. 
We set $j=L_x/4$ as the reference site to reduce the effects of the open boundary. 
Figure~\ref{fig2}(c) presents $P(r)$ for various $t_{\parallel}/J_{\perp}$ values with a fixed $J_{\rm K}$, where $E_{\rm B}/J_{\perp}$ is constant and $t_{\parallel}/E_{\rm B} \propto t_{\parallel}/J_{\perp}$. 
When $t_{\parallel}/J_{\perp}=0.2$, we indeed find the development of the pair correlation with a slow power-law decay for distance $r$, suggesting a hallmark of SC. 
As $t_{\parallel}/J_{\perp}$ increases, $P(r)$ is suppressed, whereas the power-law decay trend is maintained even at $t_{\parallel} \sim J_{\perp}$. 

Since the hardcore boson model of Eq.~\eqref{eq:hcb_model} can be mapped to the XXZ model with a certain magnetization~\cite{TMatsubara1956,MACazalilla2011}, which is integrable in one dimension, we compare $P(r)$ with the corresponding spin correlation function in the XXZ model. 
In the double Kondo lattice model, because the hardcore boson ($\hat{b}_{j}$) describes a spin tetramer, not a spin-singlet dimer, the renormalization factor $\alpha=\cos\frac{\theta}{2}$ [see Eq.~\eqref{eq:zero_hole}] is required to describe $P(r)$ of the spin-singlet ($S^c=0$) pair~\cite{SM}. 
Since $\braket{\hat{b}^{\dag}_{j+r}\hat{b}_j}$ corresponds to the transverse correlation function $C^{+-}(r)=(-1)^r\langle \hat{S}^+_{j+r}\hat{S}^-_{j} \rangle$ formulated by the spin operator $\hat{S}^{\pm}_{j}$ in the XXZ chain, we plot the renormalized $\widetilde{C}^{+-}(r)=|\alpha|^2C^{+-}(r)$ in Fig.~\ref{fig2}(c). 
Using $t_{\rm eff}$ and $V_{\rm eff}$ for the given $J_{\rm K}/J_{\perp}$, $P(r)$ approaches $\widetilde{C}^{+-}(r)$ closely as $t_{\parallel}/J_{\perp}$ decreases, revealing that the pairing state in the double Kondo lattice model connects seamlessly to the BEC-like state in the hardcore boson model at $t_{\parallel}/J_{\perp} \rightarrow 0$. 
This $t_{\parallel}/J_{\perp}$ dependence is similar to the BCS-BEC crossover demonstrated in the mixD $t$-$J$ model~\cite{HSchlomer2024,HLange_arXiv}. 
Therefore, although pair formation involves multiple interactions, the crossover from the BEC region can yield promising signatures of SC across a wide range of parameters. 
Here, we set the mediator to two localized spins; however, even when the number of localized spins is increased in each unit cell, binding energy and pair correlations still arise; see the Supplemental Material~\cite{SM}. 
Hence, our finding is not specific to the two-leg Kondo ladder. 

Importantly, the pair correlations can similarly develop regardless of the details of local spin correlations when the binding energy guarantees the presence of hole-hole pairs. 
To demonstrate this, we compare the pairing properties of $J_{\rm K} > 0$ and $J_{\rm K} < 0$. 
Since $E_{\rm B}(J_{\rm K} \rightarrow \infty)=3E_{\rm B}(J_{\rm K} \rightarrow -\infty)$ [see Fig.~\ref{fig2}(b)], we set the value of $J_{\perp}/t_{\parallel}$ on the $J_{\rm K} < 0$ side to three times the value on the $J_{\rm K} > 0$ side to make a comparison at the same level of $E_{\rm B}$. 

To clarify magnetic coupling in each unit cell, we present the site-averaged spin correlations $\bar{F}^{\mu\mu}_{\perp} = (1/L_x) \sum_j \braket{ \hat{\bm{S}}^{\mu}_{j,1} \cdot \hat{\bm{S}}^{\mu}_{j,2} }$ and $\bar{F}^{cf}_{\perp} = (1/2L_x) \sum_{j,l} \braket{ \hat{\bm{S}}^{c}_{j,l} \cdot \hat{\bm{S}}^{f}_{j,l} }$ in Figs.~\ref{fig3}(a) and \ref{fig3}(e). 
Because the Kondo spin singlet ($\bar{F}^{cf}_{\perp} \rightarrow -3/4$) is formed at $J_{\rm K}>0$ while the Kondo spin triplets ($\bar{F}^{cf}_{\perp} \rightarrow 1/4$) are formed at $J_{\rm K}<0$, the spins in each unit cell adopt qualitatively different structures on the $J_{\rm K}>0$ and $J_{\rm K}<0$ sides.

As shown in Figs.~\ref{fig3}(b) and \ref{fig3}(f), while the local spin structures are qualitatively different, the pair correlations are enhanced by $J_{\rm K}$ in both cases. 
Note that although the spin correlation $\bar{F}^{cc}_{\perp}$ approaches zero at $J_{\rm K} \rightarrow \infty$ in Fig.~\ref{fig3}(a) due to the cancelation between the $S^c=0$ and $S^c=1$ contributions in Eq.~\eqref{eq:zero_hole}, $\ket{g_{N_h=2}}=\ket{S^f=0}$ at a two-hole site promotes the development of the spin-singlet ($S^c=0$) pair correlation. 
To quantify the decaying tendency of $P(r)$, we fit it using $P(r) = A_0r^{-K_{\rm SC}}+A_1r^{-K_{\rm SC}^{\prime}}\cos(Qr +\phi)$, which is a function commonly used in two-chain systems~\cite{XLu2023,YShen2023PRB,TKaneko2024JPSJ}. 
Because $A_0 > A_1$, we regard $K_{\rm SC}$ as the decay exponent, where a smaller $K_{\rm SC}$ is more favorable for the development of correlations over longer distances. 
The estimated decay exponents are presented in Figs.~\ref{fig3}(c) and \ref{fig3}(g). 
The decay exponents on both sides exhibit similar trends as a function of $J_{\rm K}/J_{\perp}$, where $K_{\rm SC}$ drops at $|J_{\rm K}|/J_{\perp}\sim 1$ and then saturates despite the increase in $|J_{\rm K}|$. 
The values of $K_{\rm SC}$ become less than 1, suggesting the dominance of the pair correlation in the Luther-Emery liquid phase~\cite{TGiamarchi2003book,MDolfi2015,XLu2023}. 
We assess the decay exponent of the charge-density-wave correlation $K_{\rm C}$ and verify $K_{\rm C} > K_{\rm SC}$ with $K_{\rm SC}<1$ and $K_{\rm C}\times K_{\rm SC} \approx 1$ in the Supplemental Material~\cite{SM}. 

To measure how well hole-hole pairs are formed, we compute the fraction of holes forming rung pairs among all holes, $p_{\rm pair} = 2N_{\perp}^{\rm pair} / N_h$ with $N_{\perp}^{\rm pair} = \sum_j \big\langle \prod_{l}\prod_{\sigma}(1-\hat{n}^c_{j,l,\sigma}) \big\rangle$. 
Figures~\ref{fig3}(d) and \ref{fig3}(h) show the calculated $p_{\rm pair}$. 
As $|J_{\rm K}|/J_{\perp}$ increases, $p_{\rm pair}$ increases associated with the binding energy [insets of Figs.~\ref{fig3}(d) and \ref{fig3}(h)]. 
The drop of $K_{\rm SC}$ coincides with the increase of $p_{\rm pair}$, indicating that the presence of hole-hole pairs plays a crucial role in developing the pair correlations. 
Therefore, our numerical demonstrations suggest that, when the binding energy guarantees the presence of hole-hole pairs, the pair correlations can develop regardless of the details of local spin correlations in each unit cell. 

\begin{figure}[t]
\begin{center}
\includegraphics[width=\columnwidth]{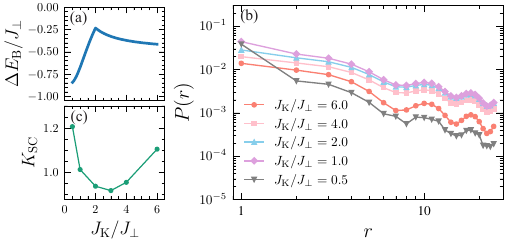} 
\caption{(a) Difference of the binding energies with and without the interchain hopping $t_{\perp}$, where $\Delta E_{\rm B}$ for $t_\perp/J_\perp=0.5$ and $t_{\parallel}=0$ is plotted. 
(b) Pair correlation function $P(r)$ and (c) decay exponent $K_{\rm SC}$ for $t_\perp/J_\perp=0.5$ and $t_{\parallel}/J_\perp=1.0$.} 
\label{fig4}
\end{center}
\end{figure}

To discuss the robustness of the above picture, we consider a factor that disrupts pair formation. 
Here, we introduce the interchain hopping term, $\hat{H}_{t_{\perp}}=-t_{\perp} \sum_{j,\sigma} ( \hat{c}^{\dag}_{j,1,\sigma} \hat{c}_{j,2,\sigma} + {\rm H.c.} )$, where $t_{\perp}$ is the hopping amplitude. 
If interchain hopping is freely allowed, $E_0(1)$ (energy of one hole) gains kinetic energy due to $t_{\perp}$, resulting in the binding energy $E_{\rm B}$ [Eq.~\eqref{eq:EB}] being smaller. 
However, this idea does not always hold in the Kondo lattice because electron mobility is affected by localized spins. 
Based on the single-hole state of Eq.~\eqref{eq:one_hole}, the hopping amplitude is $\prescript{}{1,0}{\bra{g^{\sigma}_{N_h=1}}}\hat{H}_{t_{\perp}}\ket{g^{\sigma}_{N_h=1}}_{0,1}=-t_{\perp}\cos\phi$, which vanishes at $J_{\rm K}/J_{\perp}=2$. 
The vanishing of this first-order contribution of $t_{\perp}$ strongly affects $E_{\rm B}$, which incorporates all contributions within a unit cell. 
Figure~\ref{fig4}(a) shows the change in the binding energy due to $t_{\perp}$, i.e., $\Delta E_{\rm B} = E_{\rm B}(t_{\perp}\neq0) - E_{\rm B}(t_{\perp}=0)$~\cite{SM}. 
The reduction in $E_{\rm B}$ is suppressed at $J_{\rm K}/J_{\perp}=2$. 
Therefore, pair formation can be maintained by suppressing hopping via Kondo coupling, which is a feature not present in the two-leg $t$-$J$ model. 

To numerically demonstrate the effect of $\Delta E_{\rm B}$ on the pairing state, Fig.~\ref{fig4}(b) shows the $J_{\rm K}$ dependence of $P(r)$ with $t_{\perp}$. 
We indeed find that the pair correlations increase around $J_{\rm K}/J_{\perp}=2$ and then decrease as $J_{\rm K}/J_{\perp}$ increases. 
The extracted decay exponents are presented in Fig.~\ref{fig4}(c). 
The decay exponent $K_{\rm SC}$ is minimized at an intermediate $J_{\rm K}/J_{\perp}$. 
While the minimum point is slightly shifted from $J_{\rm K}/J_{\perp}=2$ because $E_{\rm B}$ itself increases with $J_{\rm K}$, the $J_{\rm K}$ dependence of $K_{\rm SC}$ shows good agreement with that of $\Delta E_{\rm B}$. 
Therefore, screening of interchain hopping via $J_{\rm K}$ allows the presence of pairs protected by $E_{\rm B}$, which ensures a signature favorable for SC. 
This pairing arises from the insertion of localized spins and thus goes beyond the pairing mechanism in the $t$-$J$ model. 

\textit{Summary.} We show that the correlated electron systems that can be mapped onto the hardcore boson model can exhibit promising signatures of SC even when multiple spins are complexly coupled. 
Our DMRG calculations on the double Kondo lattice model demonstrate that the pair correlation develops as in the effective model, and the pairing state exhibits a crossover from the BEC regime as the parameters deviate from the strong-coupling limit. 
Moreover, we find that, when two systems have substantial and comparable binding energies, their pair correlations develop similarly, regardless of the details of the local spin correlations. 

Although the one-dimensional ladder is used in our demonstration, the same conclusion is expected for bilayers, since the mapping to the hardcore boson model is valid even in two-dimensional lattices. 
Although one might consider using the double Kondo lattice model to describe bilayer nickelate superconductors, we should pay attention to its application because both the $d_{x^2-y^2}$ and $d_{3z^2-r^2}$ orbitals are essentially involved in the Fermi surfaces of La$_3$Ni$_2$O$_7$~\cite{ZLuo2023,QGYang2023}, i.e., it is difficult to consider that either orbital is fully localized. 
If one of the two orbitals can be fully localized, the double Kondo lattice model may provide a reasonable description. 
Our discussion is not limited to a specific model because the attractive Hubbard, mixD $t$-$J$, and double Kondo lattice models can be understood within the same framework; therefore, our theory extends the guidelines for research on high-temperature SC. 

\textit{Acknowledgments.} This work was supported by Grants-in-Aid for Scientific Research from JSPS, KAKENHI Grants No.~JP24K06939, No.~JP24H00191, No.~JP24K01333, and No.~JP25H01252.
M.K. was supported by the Research Fellowship for Young Scientists, Grant No.~JP25KJ1758. 
R.U. was supported by the Program for Leading Graduate Schools: ``Interactive Materials Science Cadet Program'' and JST SPRING, Grant No.~JPMJSP2138. 
The DMRG calculations were performed using the ITENSOR library~\cite{ITensor,ITensor2}.

\textit{Data availability.} The data that support the findings of this article are openly available~\cite{data}.

\bibliography{reference}

\begin{thebibliography}{77}
\bibitem [{CGC()}]{CGC}%
  \BibitemOpen
  \href@noop {} {}\bibinfo {note} {The spin state $\ket{S_{\rm tot}, M; S^f, S^c}$ can be decomposed into $\ket{S^f,M^f}\otimes\ket{S^c,M^c}$ with the Clebsch-Gordan coefficients, where $M^{\mu}$ is the spin magnetization in orbital $\mu$.}\BibitemShut {Stop}%
\end{thebibliography}

\end{document}


\title{Supplemental Material: Superconductivity in doped spin multimer systems}
\author{Ritsuki Hirabayashi, Masataka Kakoi, Ryota Ueda, Kazuhiko Kuroki, and Tatsuya Kaneko}
\affiliation{Department of Physics, The University of Osaka, Toyonaka, Osaka 560-0043, Japan}
\date{\today}
\maketitle

\section{Analysis of the four-site model\label{sec:4_site_model}}

We analyze the low-energy structure of the four-site Kondo-Hubbard model [two itinerant-electron ($c$) sites and two localized-spin ($f$) sites], where the Hamiltonian reads
\begin{align}\label{eq:4-site_model}
\hat{H}_\perp = J_{\perp} \hat{\bm{S}}^f_{1} \cdot \hat{\bm{S}}^f_{2}
+ J_{\rm K} \sum_{l=1}^2 \hat{\bm{S}}^c_{l} \cdot \hat{\bm{S}}^f_{l} 
+ U\sum_{l=1}^2 \hat{n}^c_{l,\uparrow} \hat{n}^c_{l,\downarrow}
-t_{\perp} \sum_{\sigma} \left(\hat{c}^{\dag}_{1,\sigma} \hat{c}^{}_{2,\sigma} + {\rm H.c.}\right).
\end{align}
Here, $\hat{c}_{l,\sigma}$ ($\hat{c}^{\dag}_{l,\sigma}$) is the annihilation (creation) operator of an electron with spin $\sigma$~$(=\uparrow,\downarrow)$ on site $l$~(=1,2), and $\hat{n}_{l,\sigma}^{c}=\hat{c}^{\dag}_{l,\sigma} \hat{c}_{l,\sigma}$ and $\hat{\bm{S}}_{l}^{c}= \sum_{\alpha,\beta}{\hat{c}_{l,\alpha}^{\dagger} \bm{\sigma}_{\alpha\beta} \hat{c}_{l,\beta}}/2$ are the number and spin operators, respectively, where $\bm{\sigma}$ is the vector of Pauli matrices. 
The operator $\hat{\bm{S}}_{l}^f$ describes a localized $S=1/2$ spin. 
The parameters $J_{\perp}$, $J_{\rm K}$, $U$, and $t_{\perp}$ are the spin interaction between the localized spins, Kondo coupling, on-site repulsion, and hopping amplitude, respectively. 
Both the number operator $\hat{N}^c=\sum_{\sigma} (\hat{n}^c_{1,\sigma}+\hat{n}^c_{2,\sigma})$ and the total spin operator $\hat{\bm{S}}_{\rm tot}=\sum_{l=1}^2(\hat{\bm{S}}^f_l+\hat{\bm{S}}^c_l)$ commute with $\hat{H}_\perp$. 
Accordingly, $N^c$, $S_{\rm tot}$, and the magnetization $M$ are conserved, and the Hamiltonian can be block diagonalized into the sectors labeled by $(N^c, S_{\rm tot}, M)$. 
Table~\ref{tab:Hilbert_dimension} shows the dimensions of the Hilbert space in each $(S_{\rm tot},M)$ sector for $N^c=2$. 
The fourth column shows the dimension of the restricted Hilbert space without double occupancy of the itinerant electrons.

To construct an orthonormal basis for the Hilbert space, we introduce the state $\ket{S_{\rm tot}, M; S^f, S^c}_{n^c_1,n^c_2}$. 
Here, $S^{\mu}$ denotes the total spin of orbital $\mu$~(=$f,c$) and $n^c_l=n^c_{l,\uparrow}+n^c_{l,\downarrow}$ denotes the number of itinerant electrons at $l$.
The states are constructed from the direct-product states of the localized-spin part $\ket{S^f,M^f}$ and the itinerant-electron part $\ket{S^c,M^c}_{n^c_1,n^c_2}$ using Clebsch--Gordan coefficients, i.e., 
\begin{equation}
\ket{S_{\rm tot}, M; S^f, S^c}_{n^c_1,n^c_2} = \sum_{M^f,M^c} C^{S_{\rm tot},M}_{S^f,S^c,M^f,M^c}\ket{S^f,M^f}\otimes\ket{S^c,M^c}_{n^c_1,n^c_2}, 
\end{equation}
where $M^{\mu}$ represents the spin magnetization of orbital $\mu$. 
In this basis choice, the first ($J_{\perp}$) term of the Hamiltonian~\eqref{eq:4-site_model} becomes diagonal. 
The Clebsch--Gordan coefficients relevant to the following discussion are listed in Table~\ref{tab:cg_coef}.

\begin{table}[H]
\centering
\caption{Dimensions of the Hilbert space in the $N^c=2$ sector, classified by the total spin $S_{\rm tot}$ and magnetization $M$.
The third column shows the dimension of the full Hilbert space for each $(N^c, S_{\rm tot}, M)$ sector, while the fourth column shows the restricted dimension without double occupancy of the itinerant electrons.} 
\setlength{\tabcolsep}{8pt}
\label{tab:Hilbert_dimension}
\begin{tabular}{cccc}
\hline
\hline
$S_{\rm tot}$ & $M$ & dim. (full) & dim. (no double occupancy) \\
\hline
2 & $\pm2$, $\pm1$, $0$ & 1 & 1 \\
1 & $\pm1$, 0 & 5 & 3 \\
0 & 0      & 4 & 2 \\
\hline
\multicolumn{2}{c}{Total} & 24 & 16 \\
\hline
\hline
\end{tabular}
\end{table}

\begin{table}[t]
\begin{center}
\includegraphics[width=\columnwidth]{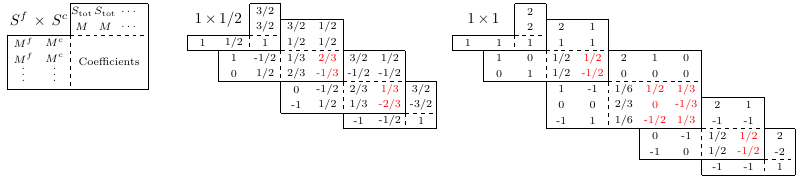} 
\caption{Clebsch--Gordan coefficients~\cite{DJGriffiths2018}, where the coefficients relevant to the present analysis are highlighted in red. 
Each entry is understood to include a square root; for example, $-1/3$ denotes $-\sqrt{1/3}$. 
Note that, for $S^f=0$ and $S^c=0$, the coefficients are $\delta_{S_{\rm tot},S^c}\delta_{M,M^c}$ and $\delta_{S_{\rm tot},S^f}\delta_{M,M^f}$, respectively.} 
\label{tab:cg_coef}
\end{center}
\end{table}

\subsection{Low-energy eigenstates when $t_{\perp}=0$}

We examine the low-energy eigenstates of the Hamiltonian~\eqref{eq:4-site_model} when $t_{\perp}=0$. 
In this case, the number operator of itinerant electrons on each site satisfies $[\hat{n}^c_l,\hat{H}_\perp]=0$, and thus $n^c_l$ is conserved. 
Consequently, in the $N^c=2$ sector, the configurations $(n^c_1, n^c_2) = (2,0), (0,2)$, and $(1,1)$ are mutually decoupled.

First, we consider the $(N^c, S_{\rm tot}, M) = (2,0,0)$ sector.
For $U \ge 0$, the ground state is in the $(n^c_1,n^c_2)=(1,1)$ sector.
We therefore consider the subspace spanned by $\{\ket{0,0; 0,0}_{1,1}, \ket{0,0; 1,1}_{1,1}\}$.
Hereafter, we omit the indices $n^c_1,n^c_2=1,1$ for simplicity.
In this basis, the Hamiltonian matrix $H^{(N^c,S_{\rm tot},M)}_\perp$ in the $(2,0,0)$ sector is given by 
\begin{equation}\label{eq:Ham_200}
    H^{(2,0,0)}_\perp = \left(\begin{matrix}
        -\frac{3J_{\perp}}{4} & -\frac{\sqrt{3}J_{\rm K}}{2}\\
         -\frac{\sqrt{3}J_{\rm K}}{2} & \frac{J_{\perp}}{4} - J_{\rm K}
    \end{matrix}\right)
    = - \left(\frac{J_{\perp}}{4} + \frac{J_{\rm K}}{2}\right) I
    -\frac{\sqrt{J_{\perp}^2 -2J_{\perp}J_{\rm K} + 4J_{\rm K}^2}}{2}
    (\sin\theta\,\sigma_x + \cos\theta\,\sigma_z),
\end{equation}
where $\theta$ satisfies
\begin{equation}
    \sin\theta = \frac{\sqrt{3}J_{\rm K}}{\sqrt{J_{\perp}^2 -2J_{\perp}J_{\rm K} + 4J_{\rm K}^2}},\quad
    \cos\theta = \frac{J_{\perp}-J_{\rm K}}{\sqrt{J_{\perp}^2 -2J_{\perp}J_{\rm K} + 4J_{\rm K}^2}}.
\end{equation}
Diagonalizing the Hamiltonian~\eqref{eq:Ham_200} yields the eigenenergies
\begin{equation}\label{eq:E_200}
    E_{\pm}^{(2,0,0)} = -\frac{J_{\perp}}{4} - \frac{J_{\rm K}}{2} \pm \frac{\sqrt{J_{\perp}^2 -2J_{\perp}J_{\rm K} + 4J_{\rm K}^2}}{2},
\end{equation}
and the corresponding eigenstates
\begin{equation}
\begin{aligned}
    \ket{\chi_+^{(2,0,0)}} &= -\sin\frac{\theta}{2}\ket{0,0;0,0} + \cos\frac{\theta}{2}\ket{0,0;1,1},\\
    \label{gs at Nh0}\ket{\chi_-^{(2,0,0)}} &= \cos\frac{\theta}{2}\ket{0,0;0,0} + \sin\frac{\theta}{2}\ket{0,0;1,1}.
\end{aligned}
\end{equation}

For the estimation of the spin gap, we consider the $S_{\rm tot}=1$ sector. 
Similarly, the Hamiltonian for $S_{\rm tot}=1$ in the basis $\{\ket{1,M; 0,1},\ket{1,M;1,0},\ket{1,M;1,1}\}$ with $M=0, \pm1$ is given by
\begin{equation}\label{eq:Ham_21M}
    H^{(2,1,M)}_\perp = \left(\begin{matrix}
        -\frac{3J_{\perp}}{4} & \frac{J_{\rm K}}{2} & 0\\
         \frac{J_{\rm K}}{2} & \frac{J_{\perp}}{4} & 0\\
         0 & 0 & \frac{J_{\perp}}{4}-\frac{J_{\rm K}}{2}
    \end{matrix}\right).
\end{equation}
The eigenenergies of the matrix~\eqref{eq:Ham_21M} are
\begin{equation}\label{eq:E_21M}
    E_{\pm}^{(2,1,M)} = -\frac{J_{\perp}}{4} \pm \frac{\sqrt{J_{\perp}^2+J_{\rm K}^2}}{2},\quad
    E_{0}^{(2,1,M)} = \frac{J_{\perp}}{4}-\frac{J_{\rm K}}{2}.
\end{equation}
For $J_{\perp}>0$, $E^{(2,1,M)}_-$ is always the first excited energy in the $N^c=2$ sector. 
The spin gap, defined as the energy difference between the $S_{\rm tot}=0$ and $S_{\rm tot}=1$ states, is therefore
\begin{equation}\label{eq:spin_gap}
    \Delta_{\rm s} := E_{-}^{(2,1,M)} - E_{-}^{(2,0,0)} = \frac{J_{\rm K}}{2} - \frac{\sqrt{J_{\perp}^2+J_{\rm K}^2}}{2} + \frac{\sqrt{J_{\perp}^2 -2J_{\perp}J_{\rm K} + 4J_{\rm K}^2}}{2}.
\end{equation}
Due to the third term that depends on the sign of $J_{\rm K}$, the spin gap $\Delta_s$ is not symmetric between the $J_{\rm K}>0$ and $J_{\rm K}<0$ sides. 

Next, we consider the $N^c=1$ sector. 
For $t_{\perp}=0$, the number of itinerant electrons at each site is conserved; therefore, the configurations $(n^c_1,n^c_2)=(1,0)$ and $(0,1)$ are decoupled.
In the basis $\big\{\ket{\frac12,M;0,\frac12}_{n^c_1,n^c_2},\ket{\frac12,M;1,\frac12}_{n^c_1,n^c_2}\big\}$, the Hamiltonian matrix $H^{(N^c,S_{\rm tot},M)}_{\perp;n^c_1,n^c_2}$ reads 
\begin{equation}\label{eq:Ham_112M}
    H^{(1,\frac12,M)}_{\perp;n^c_1,n^c_2} = \left(\begin{matrix}
        -\frac{3J_{\perp}}{4} & 0 \\
        0 & \frac{J_{\perp}}{4}-\frac{J_{\rm K}}{2}
    \end{matrix}\right)
    - \mathrm{sgn}(n^c_1-n^c_2) \left(\begin{matrix}
        0 & \frac{\sqrt{3}J_{\rm K}}{4} \\
        \frac{\sqrt{3}J_{\rm K}}{4} & 0
    \end{matrix}\right). 
\end{equation}
Here, $\mathrm{sgn}(n^c_1-n^c_2)$ in the second term arises when the Kondo coupling $(J_{\rm K})$ term is applied to the $f$-$f$ spin singlet in $\ket{\frac12,M;0,\frac12}_{n^c_1,n^c_2}$.
The eigenenergies and the eigenstates are given by
\begin{equation}\label{eq:E_112M}
    E_{\pm}^{(1,\frac12,M)} = -\frac{J_{\perp}}{4} - \frac{J_{\rm K}}{4} \pm \frac{\sqrt{J_{\perp}^2 -J_{\perp}J_{\rm K} + J_{\rm K}^2}}{2},
\end{equation}
and
\begin{equation}\label{eq:eigstate_112M}
\begin{aligned}
    \ket{\chi_+^{(1,\frac12,M)}}_{n^c_1,n^c_2} &= -\mathrm{sgn}(n^c_1-n^c_2)\sin\frac{\phi}{2}\ket{\tfrac12,M;0,\tfrac12}_{n^c_1,n^c_2} + \cos\frac{\phi}{2}\ket{\tfrac12,M;1,\tfrac12}_{n^c_1,n^c_2},\\
    \ket{\chi_-^{(1,\frac12,M)}}_{n^c_1,n^c_2} &= \cos\frac{\phi}{2}\ket{\tfrac12,M;0,\tfrac12}_{n^c_1,n^c_2} + \mathrm{sgn}(n^c_1-n^c_2)\sin\frac{\phi}{2}\ket{\tfrac12,M;1,\tfrac12}_{n^c_1,n^c_2},
\end{aligned}
\end{equation}
respectively, where $\phi$ satisfies
\begin{equation}\label{eq:def_phi}
    \sin\phi = \frac{\sqrt{3}J_{\rm K}}{2\sqrt{J_{\perp}^2 -J_{\perp}J_{\rm K} + J_{\rm K}^2}},\quad
    \cos\phi = \frac{2J_{\perp}-J_{\rm K}}{2\sqrt{J_{\perp}^2 -J_{\perp}J_{\rm K} + J_{\rm K}^2}}.
\end{equation}

For $N^c=0$, the ground state is given by the $S_{\rm tot}=0$ state, in which two localized spins form a spin singlet, with the ground-state energy $E^{(0,0,0)}=-3J_{\perp}/4$. 
Using Eqs.~\eqref{eq:E_200} and \eqref{eq:E_112M} together with $E^{(0,0,0)}$, the binding energy of two holes is given by 
\begin{align}\label{eq:binding_energy}
    E_{\rm B} &:= 2E_-^{(1,\frac12,M)} - E_-^{(2,0,0)}-E^{(0,0,0)} 
    = \frac{J_{\perp}}{2} - \sqrt{J_{\perp}^2 -J_{\perp}J_{\rm K} + J_{\rm K}^2}
    + \frac{\sqrt{J_{\perp}^2 -2J_{\perp}J_{\rm K} + 4J_{\rm K}^2}}{2}.
\end{align}

In Table~\ref{tab:low_energy_states}, we summarize the asymptotic forms of the low-energy eigenstates of the $N^c=2$ and $N^c=1$ sectors in the limits $J_{\rm K}/J_{\perp}\gg1$ and $-J_{\rm K}/J_{\perp}\gg1$. 
The fourth and fifth rows correspond to the asymptotic forms of the spin gap $\Delta_{\rm s}$ and the binding energy $E_{\rm B}$, respectively. 
The binding energy in the limits $|J_{\rm K}|/J_\perp \gg 1$ can be understood from a simple physical picture. 
When $|J_{\rm K}|/J_\perp \gg 1$, the itinerant electrons and localized spins form strong Kondo singlets ($J_{\rm K}/J_\perp \gg 1$) or triplets ($-J_{\rm K}/J_\perp \gg 1$), where the number of such $c$-$f$ pairs is equal to $N^c$. 
As a result, the second ($J_{\rm K}$) term of the Hamiltonian~\eqref{eq:4-site_model} contributes equally to $E_-^{(2,0,0)}+E^{(0,0,0)}$ and $2E_-^{(1,\frac12,M)}$ and therefore does not affect the binding energy.
Consequently, the binding energy is governed by the first ($J_{\perp}$) term, i.e., by the difference in the spin correlations of the localized spins depending on $N^c$. 
When $J_{\rm K}/J_\perp \gg 1$, the correlations $\braket{\hat{\bm{S}}^f_{1} \cdot \hat{\bm{S}}^f_{2}}$ take the values $-\frac34$, $0$, and $0$ for $N^c=0,1,$ and 2, respectively, whereas when $-J_{\rm K}/J_\perp \gg 1$, they are $-\frac34$, $-\frac12$, and $-\frac12$ (see the states listed in Table~\ref{tab:low_energy_states}). 
Hence, $E_{\rm B} \rightarrow J_{\perp} [ 2\times (0) - (0) - (-\frac34) ]=3J_{\perp}/4$ for $J_{\rm K}/J_\perp \rightarrow \infty$, while $E_{\rm B} \rightarrow J_{\perp} [ 2\times (-\frac12) - (-\frac12) - (-\frac34) ]=J_{\perp}/4$ for $J_{\rm K}/J_\perp \rightarrow -\infty$. 
This explains why forming a hole-hole pair when $J_{\rm K}/J_\perp \gg 1$ yields an additional energy gain of $J_{\perp}/2$ compared to the case of $-J_{\rm K}/J_\perp \gg 1$.

\begin{table}[H]
\centering
\caption{Summary of the low-energy eigenstates when $t_{\perp}=0$. 
In the $N^c=1$ sector, the results for $(n_1^c,n_2^c)=(1,0)$ are presented. 
The fourth and fifth rows show the asymptotic forms of the spin gap $\Delta_{\rm s}$ and the binding energy $E_{\rm B}$ at half filling, respectively.}
\setlength{\tabcolsep}{8pt}
\renewcommand{\arraystretch}{2.4}
\label{tab:low_energy_states}
\begin{tabular}{cccc}
\hline
\hline
 \noalign{\vskip -8pt} & $J_{\rm K}/J_{\perp}\gg1$ & $-J_{\rm K}/J_{\perp}\gg1$ & degeneracy\\[-3pt]
\hline
$N^c=2$, $S_{\rm tot}=0$ & $\displaystyle \frac12\ket{0,0;0,0} + \frac{\sqrt{3}}{2}\ket{0,0;1,1}$ 
& $\displaystyle \frac{\sqrt{3}}{2}\ket{0,0;0,0} - \frac12\ket{0,0;1,1}$ & 1 \\
$N^c=2$, $S_{\rm tot}=1$ & $\displaystyle \frac{1}{\sqrt{2}}\ket{1,M;0,1}-\frac{1}{\sqrt{2}}\ket{1,M;1,0}$
& $\displaystyle \frac{1}{\sqrt{2}}\ket{1,M;0,1} + \frac{1}{\sqrt{2}}\ket{1,M;1,0}$ & 3 \\
$N^c=1$, $\displaystyle S_{\rm tot}=\frac12$ & $\displaystyle \frac12\ket{\tfrac12,M;0,\tfrac12}_{1,0} + \frac{\sqrt{3}}{2}\ket{\tfrac12,M;1,\tfrac12}_{1,0}$
& $\displaystyle \frac{\sqrt{3}}{2}\ket{\tfrac12,M;0,\tfrac12}_{1,0} - \frac12\ket{\tfrac12,M;1,\tfrac12}_{1,0}$ & 4 \\
\hline
$\Delta_{\rm s}$ [Eq.~\eqref{eq:spin_gap}]
& $\displaystyle J_{\rm K} - \frac{J_{\perp}}{4}$ & $\displaystyle \frac{J_{\perp}}{4}+\frac{5J_{\perp}^2}{32J_{\rm K}}$ \\
$E_{\rm B}$ [Eq.~\eqref{eq:binding_energy}]
& $\displaystyle \frac{3J_{\perp}}{4}-\frac{9J_{\perp}^2}{32J_{\rm K}}$ 
& $\displaystyle \frac{J_{\perp}}{4}+\frac{9J_{\perp}^2}{32J_{\rm K}}$ \\
\hline
\hline
\end{tabular}
\end{table}

\subsection{Effect of $t_{\perp}$ on the binding energy}

We examine how the low-energy structure is modified by $t_{\perp}\neq0$, where $n^c_l$ is no longer conserved.
In the $N^c=1$ sector, the Hamiltonian~\eqref{eq:4-site_model} has nonzero matrix elements between the configurations $(n^c_1,n^c_2)=(1,0)$ and $(0,1)$, thereby lifting their degeneracy, while the degeneracy with respect to $M$ is preserved.
For $S_{\rm tot}=1/2$, the Hamiltonian can be decomposed into two $2\times2$ blocks in the basis $\{\ket{\psi_{+}^{0}},\ket{\psi_{-}^{1}},\ket{\psi_{-}^{0}},\ket{\psi_{+}^{1}}\}$, where $\ket{\psi_{\pm}^{S^f}}:=\frac{1}{\sqrt{2}}\big(\ket{\frac12,M;S^f,\frac12}_{1,0}\pm\ket{\frac12,M;S^f,\frac12}_{0,1}\big)$, as 
\begin{equation}
    H^{(1,\frac12,M)}_\perp = \left(\begin{matrix}
        -\frac{3J_{\perp}}{4}-t_{\perp} & -\frac{\sqrt{3}J_{\rm K}}{4} & 0 & 0\\
        -\frac{\sqrt{3}J_{\rm K}}{4} & \frac{J_{\perp}}{4}-\frac{J_{\rm K}}{2}+t_{\perp} & 0 & 0\\
        0 & 0 & -\frac{3J_{\perp}}{4}+t_{\perp} & -\frac{\sqrt{3}J_{\rm K}}{4} \\
        0 & 0 & -\frac{\sqrt{3}J_{\rm K}}{4} & \frac{J_{\perp}}{4}-\frac{J_{\rm K}}{2}-t_{\perp}
    \end{matrix}\right).
\end{equation}
From the upper-left block, we obtain the eigenenergies
\begin{equation}\label{eq:E_(1,1/2,M)+}
    E_{1,\pm}^{(1,\frac12,M)}=-\frac{J_{\perp}}{4} - \frac{J_{\rm K}}{4} \pm \frac12\sqrt{\left(J_{\perp} -\frac{J_{\rm K}}{2} + 2t_{\perp}\right)^{\!2} + \frac34J_{\rm K}^2},
\end{equation}
while from the lower-right block, we obtain
\begin{equation}\label{eq:E_(1,1/2,M)-}
    E_{2,\pm}^{(1,\frac12,M)}=-\frac{J_{\perp}}{4} - \frac{J_{\rm K}}{4} \pm \frac12\sqrt{\left(J_{\perp} -\frac{J_{\rm K}}{2} - 2t_{\perp}\right)^{\!2} + \frac34J_{\rm K}^2}.
\end{equation}
The ground state changes at $J_{\rm K}=2J_{\perp}$; when $t_{\perp}>0$, $E_{2,-}^{(1,\frac12,M)}<E_{1,-}^{(1,\frac12,M)}$ for $J_{\rm K}>2J_{\perp}$, while the inequality is reversed for $J_{\rm K}<2J_{\perp}$.

In the $N^c=2$ sector, nonzero matrix elements exist between the configurations $(n^c_1, n^c_2) = (2,0), (0,2)$, and $(1,1)$.
In the $S_{\rm tot}=0$ sector, using the basis $\{\ket{\chi_+^{(2,0,0)}}, \ket{\chi_-^{(2,0,0)}}, \ket{D}_+, \ket{D}_-\}$, where $\ket{D}_{\pm} := (\ket{0,0;0,0}_{2,0}\pm\ket{0,0;0,0}_{0,2})/\sqrt{2}$ are appended, we have 
\begin{equation}\label{eq:N=2_S=0_t_perp_Hamiltonian}
    H^{(2,0,0)}_\perp = \left(\begin{matrix}
        E_+^{(2,0,0)} & 0 & 2t_{\perp}\sin\frac{\theta}{2} & 0\\
        0 & E_-^{(2,0,0)} & -2t_{\perp}\cos\frac{\theta}{2} &0\\
        2t_{\perp}\sin\frac{\theta}{2} & -2t_{\perp}\cos\frac{\theta}{2} & U - \frac{3J_{\perp}}{4} & 0 \\
        0 & 0 & 0 & U - \frac{3J_{\perp}}{4}
    \end{matrix}\right).
\end{equation}
Diagonalizing the $3\times3$ block in Eq.~\eqref{eq:N=2_S=0_t_perp_Hamiltonian}, one obtains the lowest-energy state in the $N^c=2$ sector and then the analytic form of the binding energy with Eq.~\eqref{eq:E_(1,1/2,M)+} or \eqref{eq:E_(1,1/2,M)-}. 
The expression of $E_{\rm B}$ is complicated, and we do not write it explicitly here. 
Instead, for $U\gg J_{\perp}$, the difference in the binding energy from that at $t_{\perp}=0$, approximated up to the second order in $t_{\perp}$, is given by 
\begin{equation}\label{eq:Delta_Eb}
    \Delta E_{\rm B} := E_{\rm B}(t_{\perp}\neq0) - E_{\rm B}(t_{\perp}=0) 
    \simeq -2t_{\perp}|\cos\phi| - \frac{2t_{\perp}^2\sin^2\phi}{\sqrt{J_{\perp}^2 -J_{\perp}J_{\rm K} + J_{\rm K}^2}} 
    + \frac{4t_\perp^2\cos^2\frac{\theta}{2}}{U-\frac{3J_{\perp}}{4}-E_-^{(2,0,0)}}.
\end{equation}
The hopping $t_{\perp}$ reduces the lowest energy at $N^c=1$, which appears as the first and second terms in Eq.~\eqref{eq:Delta_Eb}.
The hopping $t_{\perp}$ also reduces the energy at $N^c=2$ via the exchange process, which increases the binding energy as the third term. 
Including both contributions, the binding energy typically decreases compared to the case at $t_{\perp}=0$. 

The first term in Eq.~\eqref{eq:Delta_Eb} (first order in $t_{\perp}$) vanishes at $\cos\phi=0$, which is equivalent to $J_{\rm K}=2J_{\perp}$ [see Eq.~\eqref{eq:def_phi}].
This can be intuitively understood as follows. 
For $t_{\perp}=0$, the lowest-energy state at $N^c=1$ is two-fold degenerate, spanned by $\ket{\chi_-^{(1,\frac12,M)}}_{1,0}$ and $\ket{\chi_-^{(1,\frac12,M)}}_{0,1}$. 
Using Eq.~\eqref{eq:eigstate_112M}, the matrix element of the hopping term in Eq.~\eqref{eq:4-site_model} (denoted by $\hat{H}_{t_{\perp}}$) between these states is given by 
\begin{equation}
    \prescript{}{1,0}{\bra{\chi_-^{(1,\frac12,M)}}}
    \hat{H}_{t_{\perp}}
    \ket{\chi_-^{(1,\frac12,M)}}_{0,1}
    = -t_{\perp}\left(\cos^2\frac{\phi}{2} - \sin^2\frac{\phi}{2}\right) = -t_{\perp}\cos\phi.
\end{equation}
At $J_{\rm K}=2J_{\perp}$, this matrix element is zero, implying that the hopping between $\ket{\chi_-^{(1,\frac12,M)}}_{1,0}$ and $\ket{\chi_-^{(1,\frac12,M)}}_{0,1}$ is prohibited.
This means that the energy change in the presence of $t_{\perp}$ arises only from the second-order processes via the excited state $\ket{\chi_+^{(1,\frac12,M)}}$ in the $N^c=1$ sector and the exchange processes in the $N^c=2$ sector. 
For $J_{\rm K}\simeq 2J_{\perp}$, the reduction of the binding energy is suppressed; therefore, the hole-hole pair is not destabilized, and superconductivity may not be adversely affected even with $t_{\perp} \ne 0$.

\section{Effective hardcore boson model for the double Kondo lattice model}

In doped $t$-$J$ ladder or bilayer models with a strong rung exchange $J_{\perp}$, a picture in which tightly bound hole-hole pairs move in a spin-singlet background is commonly adopted. 
Consequently, these systems can be mapped onto effective models by treating hole-hole pairs (or spin singlets) as hardcore bosons~\cite{MTroyer1996}. 
As shown in Sec.~\ref{sec:4_site_model}, the four-site Kondo-Hubbard model~\eqref{eq:4-site_model} at half-filling exhibits a nonzero binding energy for $J_{\rm K}\neq 0$ and $J_{\perp}>0$. 
This result allows the doped double Kondo lattice model with the hopping term along the chain (or in-plane) direction given by
\begin{equation}\label{eq:hopterm}
    \hat{H}_{\parallel} = -t_{\parallel} \sum_{\braket{i,j}}\sum_{l} \sum_{\sigma}
     \left(\hat{c}^{\dag}_{i,l,\sigma} \hat{c}^{}_{j,l,\sigma} + {\rm H.c.}\right)
\end{equation}
to be mapped onto an effective boson model, as we demonstrate below.

\subsection{Mapping to the hardcore boson $t$-$V$ model \label{sec:mapping}}

We define $\ket{g^{\rm (T)}}$ as the ground state of the $(N^c,S_{\rm tot})=(2,0)$ sector, in which two itinerant electrons and two localized spins form a {\it tetramer}. 
For example, $\ket{g^{\rm (T)}} = \ket{\chi_-^{(2,0,0)}}$ for $t_{\perp}=0$.
When each rung is completely decoupled ($t_{\parallel}=0$), at half filling, the ground state is $\ket{\Psi_0}=\ket{g^{\rm (T)}}\otimes\ket{g^{\rm (T)}}\otimes\cdots\otimes\ket{g^{\rm (T)}}$. 
When $N_h= 2n$ holes are doped, the ground state is given by any linear combination of the states
\begin{equation}
\ket{j_1,j_2,...,j_{n}} = \left(\prod_{k=1}^{n} \hat{b}^{\dag}_{j_k}\right)\ket{\Psi_0},
\end{equation}
where $\hat{b}^{\dag} := \ket{\rm hh}\bra{g^{\rm (T)}}$ creates a two-hole bound state $\ket{\rm hh} = \ket{0,0;0,0}_{0,0}$.
Even for $t_{\parallel} \ne 0$, when $t_{\parallel}\ll \Delta_{\rm s},E_{\rm B}(<J_{\perp})$, the effective model obtained by projecting the Hamiltonian onto the subspace
\begin{equation}\label{eq:subspace}
\mathcal{H}_n = \mathrm{span}\{\ket{j_1,j_2,...,j_{n}}\} 
\end{equation}
should provide a good description of the low-energy physics. 
When the in-chain (in-plane) hopping term $\hat{H}_{\parallel}$ acts between a tetramer state $\ket{g^{\rm (T)}}$ and a two-hole bound state $\ket{\rm hh}$, two {\it trimers} are excited. 
Second-order perturbation processes involving these trimer states as intermediate states give rise to both the hopping of a boson (i.e., two-hole bound state) and interaction between two bosons.
Similarly, when $\hat{H}_{\parallel}$ acts between neighboring tetramer ($N^c=2$) states, one $N^c=1$ state and one $N^c=3$ state are excited. 
Second-order perturbation processes through these intermediate states give rise to interactions between bosons. 
To describe the effective model, we use the relations
\begin{equation}\label{eq:convert_boson_ops}
\begin{gathered}
    \hat{b}^{\dag}_{i}\hat{b}^{}_{j} 
    = \big(\ket{\rm hh}\bra{g^{\rm (T)}}\big)_i
    \otimes\big(\ket{g^{\rm (T)}}\bra{\rm hh}\big)_j, \quad
    (1-\hat{n}^{b}_{i})\hat{n}^{b}_{j}
    = \big(\ket{g^{\rm (T)}}\bra{g^{\rm (T)}}\big)_i
    \otimes\big(\ket{\rm hh}\bra{\rm hh}\big)_j,
    \\[1mm]
    (1-\hat{n}^{b}_{i})(1-\hat{n}^{b}_{j})
    = \big(\ket{g^{\rm (T)}}\bra{g^{\rm (T)}}\big)_i
    \otimes\big(\ket{g^{\rm (T)}}\bra{g^{\rm (T)}}\big)_j
\end{gathered} 
\end{equation}
with the number operator $\hat{n}^b_j=\hat{b}^{\dag}_j\hat{b}_j$. 
The low-energy effective Hamiltonian for the two-hole bound states can be mapped, including all second-order processes, onto the following $t$-$V$ model:
\begin{equation}\label{eq:t-V_model}
    \hat{H}_{\rm eff} = -t_{\rm eff} \sum_{\braket{i,j}} \left(\hat{b}^{\dag}_{i}\hat{b}^{}_{j} + {\rm H.c.}\right)
    + V_{\rm eff} \sum_{\braket{i,j}} \hat{n}^b_i\hat{n}_j^b
    + \epsilon_{\rm eff} \sum_j \hat{n}_j^b.
\end{equation}
Note that this expression assumes periodic boundary conditions. 
The large-$J_{\perp}$ $t$-$J$ ladder or bilayer models can also be mapped onto the same form~\cite{MTroyer1996}. 
In the limit $J_{\parallel} = t_{\perp}=0$ in such $t$-$J$ models, the parameters satisfy $V_{\rm eff}=2t_{\rm eff}=4t_{\parallel}^2/J_{\perp}$.

Here, we derive the explicit expressions of $t_{\rm eff}$, $V_{\rm eff}$, and $\epsilon_{\rm eff}$ as functions of $t_{\parallel}$, $J_{\perp}$, and $J_{\rm K}$ for the simplest case at $t_{\perp}=0$, where $\ket{g^{\rm (T)}} = \ket{\chi_-^{(2,0,0)}}$. 
Since the hopping $t_{\parallel}$ acts only between nearest-neighbor rungs, where the Hamiltonian can be written as $\hat{H} = \sum_{\braket{i,j}} \hat{H}_{ij}$, and the projection onto the low-energy subspace $\mathcal{H}_n$ excludes processes involving three or more rungs, the contributions in second-order perturbation processes are restricted to those only involving two nearest-neighbor rungs. 
Therefore, in the following discussion, we consider only two rungs. 
We consider all second-order perturbation processes between a tetramer and a neighboring two-hole bound state, as well as those between neighboring tetramers.
The former involves 16 possible intermediate states, in which one rung state satisfies $(N^c,S_{\rm tot},M)=(1,\tfrac12,\frac{\sigma}{2})$, and the other satisfies $(1,\tfrac12,-\frac{\sigma}{2})$:
\begin{align}\label{eq:2-order-perturb}
    \ket{\rm hh}\otimes\ket{\chi_-^{(2,0,0)}} & \to \left\{\begin{matrix}
        \ket{\psi_1^{\sigma\bar{\sigma}}} = 
        \ket{\chi_-^{(1,\frac{1}{2},\frac{\sigma}{2})}}_{1,0}\otimes\ket{\chi_-^{(1,\frac{1}{2},-\frac{\sigma}{2})}}_{0,1},\quad
        \ket{\psi_2^{\sigma\bar{\sigma}}} = 
        \ket{\chi_-^{(1,\frac{1}{2},\frac{\sigma}{2})}}_{0,1}\otimes\ket{\chi_-^{(1,\frac{1}{2},-\frac{\sigma}{2})}}_{1,0} \\[2mm]
        \ket{\psi_3^{\sigma\bar{\sigma}}} = 
        \ket{\chi_+^{(1,\frac{1}{2},\frac{\sigma}{2})}}_{1,0}\otimes\ket{\chi_+^{(1,\frac{1}{2},-\frac{\sigma}{2})}}_{0,1},\quad
        \ket{\psi_4^{\sigma\bar{\sigma}}} = 
        \ket{\chi_+^{(1,\frac{1}{2},\frac{\sigma}{2})}}_{0,1}\otimes\ket{\chi_+^{(1,\frac{1}{2},-\frac{\sigma}{2})}}_{1,0} \\[2mm]
        \ket{\psi_5^{\sigma\bar{\sigma}}} = 
        \ket{\chi_+^{(1,\frac{1}{2},\frac{\sigma}{2})}}_{1,0}\otimes\ket{\chi_-^{(1,\frac{1}{2},-\frac{\sigma}{2})}}_{0,1},\quad
        \ket{\psi_6^{\sigma\bar{\sigma}}} = 
        \ket{\chi_+^{(1,\frac{1}{2},\frac{\sigma}{2})}}_{0,1}\otimes\ket{\chi_-^{(1,\frac{1}{2},-\frac{\sigma}{2})}}_{1,0} \\[2mm]
        \ket{\psi_7^{\sigma\bar{\sigma}}} =
        \ket{\chi_-^{(1,\frac{1}{2},\frac{\sigma}{2})}}_{1,0}\otimes\ket{\chi_+^{(1,\frac{1}{2},-\frac{\sigma}{2})}}_{0,1},\quad
        \ket{\psi_8^{\sigma\bar{\sigma}}} = 
        \ket{\chi_-^{(1,\frac{1}{2},\frac{\sigma}{2})}}_{0,1}\otimes\ket{\chi_+^{(1,\frac{1}{2},-\frac{\sigma}{2})}}_{1,0}
    \end{matrix}\right\} \notag\\
    &\to \left\{\begin{matrix}
        \ket{\rm hh}\otimes\ket{\chi_-^{(2,0,0)}}\\[2mm]
        \ket{\chi_-^{(2,0,0)}}\otimes\ket{\rm hh}
    \end{matrix}\right\} .
\end{align}
From the selection rule of angular momentum, intermediate states containing an $S_{\rm tot}=3/2$ state do not contribute. 
To see this, note that $\hat{c}^{}_{l,\sigma}$ corresponds to a spherical tensor operator of rank $1/2$ in spin space. 
By the Wigner--Eckart theorem, a necessary condition for $\bra{S'_{\rm tot},M'}\hat{c}^{}_{l,\sigma}\ket{S_{\rm tot},M}$ to be nonzero is $|S_{\rm tot} - 1/2| \leq S_{\rm tot}' \leq S_{\rm tot} + 1/2$. 
Since the transition $S_{\rm tot}=0\to S'_{\rm tot}=3/2$ does not satisfy this condition, its matrix element vanishes identically. 
The energy difference $\Delta E_k$ between the intermediate state $\ket{\psi_k^{\sigma\bar{\sigma}}}$ and the initial (final) state given by
\begin{equation}\label{eq:def_DE}
    \Delta E_k = \left\{\begin{array}{ll}
        E_{\rm B} & \text{for $k=1,2$} \\
        E_{\rm B} + 2D & \text{for $k=3,4$} \\
        E_{\rm B} + D & \text{for $k=5,6,7,8$}
    \end{array}\right.
    ,
\end{equation}
where
\begin{equation}\label{eq:def_D}
    D = E_+^{(1,\frac12,M)} - E_-^{(1,\frac12,M)} = \sqrt{J_{\perp}^2 -J_{\perp}J_{\rm K} + J_{\rm K}^2}.
\end{equation}

For the perturbation processes between neighboring tetramers, an $N^c=3$ state is included in the intermediate state.
Before listing all intermediate states, we consider the eigenstates in the $(N^c,S_{\rm tot},M)=(3,\frac12,M)$ sector. 
For an arbitrary $t_{\perp}$, the operator
\begin{equation}
\hat{F} = \hat{c}^{}_{1,\uparrow}\hat{c}^{}_{1,\downarrow} - \hat{c}^{}_{2,\uparrow}\hat{c}^{}_{2,\downarrow},
\end{equation}
satisfies
\begin{equation}
[\hat{H}_\perp,\hat{F}] = -U\hat{F},\qquad
[\hat{H}_\perp,\hat{F}^{\dag}] = U\hat{F}^{\dag}.
\end{equation}
Therefore, if $\ket{\chi}$ is an eigenstate of $\hat{H}_\perp$ with eigenvalue $E$, $\hat{F}\ket{\chi}(\neq0)$ and $\hat{F}^{\dagger}\ket{\chi}(\neq0)$ are also eigenstates with eigenvalues $E-U$ and $E+U$, respectively.
Using this relation, the eigenstates in the $(N^c,S_{\rm tot},M)=(3,\frac12,M)$ sector can be constructed from those in the $(1,\frac12,M)$ sector as
\begin{equation}\label{eq:chi_312M}
\ket{\chi^{(3,\frac12,M)}_{\pm}}_{n^c_1,n^c_2} = \hat{F}^{\dagger}\ket{\chi^{(1,\frac12,M)}_{\pm}}_{2-n^c_1,2-n^c_2},
\end{equation}
where $n^c_1+n^c_2=3$. 
The corresponding eigenvalues are
\begin{equation}
E^{(3,\frac12,M)}_{\pm} = E^{(1,\frac12,M)}_{\pm} + U.
\end{equation}
For convenience, we define
\begin{equation}\label{eq:def_varphi}
    \ket{^1\varphi_k^{\sigma\bar{\sigma}}} = \hat{F}^{\dagger}\otimes\hat{1}\ket{\psi_k^{\sigma\bar{\sigma}}},\qquad
    \ket{^2\varphi_k^{\sigma\bar{\sigma}}} = \hat{1}\otimes\hat{F}^{\dagger}\ket{\psi_k^{\sigma\bar{\sigma}}}\qquad
    \text{for $k=1,2,\cdots,8$.}
\end{equation}
The perturbation processes include 32 possible intermediate states:
\begin{align}
    \ket{\chi_-^{(2,0,0)}}\otimes\ket{\chi_-^{(2,0,0)}} & \to \Big\{
    \ket{^1\varphi_1^{\sigma\bar{\sigma}}},\ \cdots,\ \ket{^1\varphi_8^{\sigma\bar{\sigma}}},\ 
    \ket{^2\varphi_1^{\sigma\bar{\sigma}}},\ \cdots,\ \ket{^2\varphi_8^{\sigma\bar{\sigma}}}
    \Big\} \to \ket{\chi_-^{(2,0,0)}}\otimes\ket{\chi_-^{(2,0,0)}}.
\end{align}
As in process \eqref{eq:2-order-perturb}, intermediate states containing an $S_{\rm tot}=3/2$ state are excluded.
The energy difference $\Delta E_k'$ between the intermediate state $\ket{^1\varphi_k^{\sigma\bar{\sigma}}}$ ($\ket{^2\varphi_k^{\sigma\bar{\sigma}}}$) and the initial (final) state is given by
\begin{equation}\label{eq:def_DE_prime}
    \Delta E_k' = \left\{\begin{array}{ll}
        U + 2D' & \text{for $k=1,2$} \\
        U + 2D' + 2D & \text{for $k=3,4$} \\
        U + 2D' + D & \text{for $k=5,6,7,8$}
    \end{array}\right.
    ,
\end{equation}
where $D$ is presented in Eq.~\eqref{eq:def_D} and 
\begin{equation}
    D' = E_-^{(1,\frac12,M)} - E_-^{(2,0,0)} = \frac{J_{\rm K}}{4} -\frac{\sqrt{J_{\perp}^2 -J_{\perp}J_{\rm K} + J_{\rm K}^2}}{2} + \frac{\sqrt{J_{\perp}^2 -2J_{\perp}J_{\rm K} + 4J_{\rm K}^2}}{2}.
\end{equation} 

We define $\hat{P}^{(2)}$, $\hat{Q}^{(2)}_1$ and $\hat{Q}^{(2)}_2$ as the projection operators onto the subspaces spanned by 
\begin{gather}
\left\{\ket{\rm hh}\otimes\ket{\rm hh},\ \ket{\rm hh}\otimes\ket{\chi_-^{(2,0,0)}},\ \ket{\chi_-^{(2,0,0)}}\otimes\ket{\rm hh},\ \ket{\chi_-^{(2,0,0)}}\otimes\ket{\chi_-^{(2,0,0)}}\right\},
\notag \\
\left\{\ket{\psi_1^{\uparrow\downarrow}},\,\ket{\psi_1^{\downarrow\uparrow}},\,\ket{\psi_2^{\uparrow\downarrow}},\,\cdots,\,\ket{\psi_8^{\downarrow\uparrow}}\right\},
\notag 
\end{gather} 
and 
\begin{equation}
\left\{\ket{^1\varphi_1^{\uparrow\downarrow}},\,\cdots,\,\ket{^1\varphi_8^{\downarrow\uparrow}},\,\ket{^2\varphi_1^{\uparrow\downarrow}},\,\cdots,\,\ket{^2\varphi_8^{\downarrow\uparrow}}\right\},
\notag 
\end{equation} 
respectively.
The effective Hamiltonian involving all second-order perturbation processes is therefore
\begin{align}\label{eq:effective_model_2nd_order}
    \hat{H}_{\rm eff} &= \sum_{\braket{i,j}} \hat{P}^{(2)}\hat{H}_{ij}\hat{P}^{(2)}- \sum_{\braket{i,j}} 
    \hat{P}^{(2)}\hat{H}_{ij}\hat{Q}^{(2)}_1
    \left(\sum_{k,\sigma}\frac{1}{\Delta E_k}\ket{\psi_k^{\sigma\bar{\sigma}}}\bra{\psi_k^{\sigma\bar{\sigma}}}\right)
    \hat{Q}^{(2)}_1\hat{H}_{ij}\hat{P}^{(2)} \notag\\
    &-\sum_{\braket{i,j}} 
    \hat{P}^{(2)}\hat{H}_{ij}\hat{Q}^{(2)}_2
    \left(\sum_{\nu,k,\sigma}\frac{1}{\Delta E_k'}\ket{^{\nu\!}\varphi_k^{\sigma\bar{\sigma}}}\bra{^{\nu\!}\varphi_k^{\sigma\bar{\sigma}}}\right)
    \hat{Q}^{(2)}_2\hat{H}_{ij}\hat{P}^{(2)}, 
\end{align}
where $\Delta E_k$ and $\Delta E_k'$ are defined in Eqs.~\eqref{eq:def_DE} and \eqref{eq:def_DE_prime}. 
The hopping $\hat{P}^{(2)}\hat{H}_{ij}\hat{Q}^{(2)}_1$ can be evaluated using
\begin{equation}\label{eq:hopelem}
\begin{gathered}
    \bra{\chi_-^{(2,0,0)}}\hat{c}^{\dag}_{1,\sigma}
    \ket{\chi_-^{(1,\frac12,-\frac{\sigma}{2})}}_{0,1}
    = \bra{\chi_-^{(2,0,0)}}\hat{c}^{\dag}_{2,\sigma}
    \ket{\chi_-^{(1,\frac12,-\frac{\sigma}{2})}}_{1,0}
    = \frac{\sigma}{\sqrt{2}}\cos\!\left(\frac{\theta-\phi}{2}\right),\\[1mm]
    \bra{\chi_-^{(2,0,0)}}\hat{c}^{\dag}_{1,\sigma}
    \ket{\chi_+^{(1,\frac12,-\frac{\sigma}{2})}}_{0,1}
    = -\bra{\chi_-^{(2,0,0)}}\hat{c}^{\dag}_{2,\sigma}
    \ket{\chi_+^{(1,\frac12,-\frac{\sigma}{2})}}_{1,0}
    = -\frac{\sigma}{\sqrt{2}}\sin\!\left(\frac{\theta-\phi}{2}\right),\\[1mm]
    \prescript{}{1,0}{\bra{\chi_-^{(1,\frac12,\frac{\sigma}{2})}}}
    \hat{c}^{\dag}_{1,\sigma}\ket{\rm hh} 
    = \prescript{}{0,1}{\bra{\chi_-^{(1,\frac12,\frac{\sigma}{2})}}}
    \hat{c}^{\dag}_{2,\sigma}\ket{\rm hh} 
    = \cos\frac{\phi}{2},\\[1mm]
    \prescript{}{1,0}{\bra{\chi_+^{(1,\frac12,\frac{\sigma}{2})}}}
    \hat{c}^{\dag}_{1,\sigma}\ket{\rm hh} 
    = -\prescript{}{0,1}{\bra{\chi_+^{(1,\frac12,\frac{\sigma}{2})}}}
    \hat{c}^{\dag}_{2,\sigma}\ket{\rm hh}
    = -\sin\frac{\phi}{2}.
\end{gathered}
\end{equation}
Similarly, $\hat{P}^{(2)}\hat{H}_{ij}\hat{Q}^{(2)}_2$ can be evaluated by Eqs.~\eqref{eq:chi_312M}, \eqref{eq:hopelem}, and $[\hat{c}_{l,\sigma},\hat{F}^{\dag}] = (-1)^{l}\sigma \hat{c}^{\dag}_{l,\bar{\sigma}}$; e.g., $\bra{\chi_-^{(2,0,0)}}\hat{c}_{1,\sigma}\ket{\chi_-^{(3,\frac12,\frac{\sigma}{2})}}_{2,1}
= -\sigma \bra{\chi_-^{(2,0,0)}}\hat{c}^{\dag}_{1,\bar{\sigma}}\ket{\chi_-^{(1,\frac12,\frac{\sigma}{2})}}_{0,1}$. 
Using Eq.~\eqref{eq:convert_boson_ops} to simplify the effective Hamiltonian~\eqref{eq:effective_model_2nd_order}, we obtain Eq.~\eqref{eq:t-V_model}, in which $t_{\rm eff}$, $V_{\rm eff}$, and $\epsilon_{\rm eff}$ for the double Kondo lattice model are given by 
\begin{equation}\label{eq:bosonparams}
\begin{aligned}
    t_{\rm eff} &= \frac{K_{--}+K_{++}+K_{+-}^{\text{(off-diag)}}}{2},\\
    V_{\rm eff} &= K_{--}+K_{++}+K_{+-}^{\text{(diag)}} - \left( K_{--}^U+K_{++}^U+K_{+-}^U \right),\\
    \epsilon_{\rm eff} &= -\frac{K_{--}+K_{++}+K_{+-}^{\text{(diag)}}}{2}z + \left(K_{--}^U+K_{++}^U+K_{+-}^U\right)z + E_-^{(0,0,0)} - E_-^{(2,0,0)},
\end{aligned}
\end{equation}
where $z$ is the coordination number ($z=2$ for the ladder model) and 
\begin{equation}
\begin{gathered}
K_{--} = \frac{4t_{\parallel}^2}{E_{\rm B}}\cos^2\frac{\phi}{2}\cos^2\left(\frac{\theta-\phi}{2}\right), \quad 
K_{++} = \frac{4t_{\parallel}^2}{{E_{\rm B} + 2D}}\sin^2\frac{\phi}{2}\sin^2\left(\frac{\theta-\phi}{2}\right), \quad \\[1mm]
K_{+-}^{\text{(off-diag)}} = -\frac{2t_{\parallel}^2}{{E_{\rm B} + D}}\sin\phi\sin(\theta-\phi), \quad \\[1mm]
K_{+-}^{\text{(diag)}} = \frac{4t_{\parallel}^2}{{E_{\rm B} + D}}\left[\sin^2\frac{\phi}{2}\cos^2\left(\frac{\theta-\phi}{2}\right)+\cos^2\frac{\phi}{2}\sin^2\left(\frac{\theta-\phi}{2}\right) \right], \\[1mm]
K_{--}^U = \frac{2t_{\parallel}^2}{U+2D'}\cos^4\left(\frac{\theta-\phi}{2}\right), \quad 
K_{++}^U = \frac{2t_{\parallel}^2}{U+2D'+2D}\sin^4\left(\frac{\theta-\phi}{2}\right), \quad \\[1mm]
K_{+-}^U = \frac{4t_{\parallel}^2}{U+2D'+D}\cos^2\left(\frac{\theta-\phi}{2}\right)\sin^2\left(\frac{\theta-\phi}{2}\right).
\end{gathered}
\end{equation}
Note that $\epsilon_{\rm eff}$ in Eq.~\eqref{eq:bosonparams} is the expression under periodic boundary conditions. 
Here, $K_{+-}^{\text{(diag)}}$ ($K_{+-}^{\text{(off-diag)}}$) stems from the diagonal (off-diagonal) elements of the operators, which result in the $V_{\rm eff}$ ($t_{\rm eff}$) term, inside the summation in the second term of Eq.~\eqref{eq:effective_model_2nd_order}. 
The processes that pass through $\ket{\psi_k^{\sigma\bar{\sigma}}}$ ($k=1,2,3,4$) contribute equally to both the diagonal and off-diagonal elements.
Equation (6) in the main text presents $t_{\rm eff}$ and $V_{\rm eff}$ given by the leading contribution $K_{--}$.

\subsection{Correspondence with the XXZ model} 

The hardcore boson $t$-$V$ model~\eqref{eq:t-V_model} is equivalent to the XXZ model under the transformations $\hat{b}^{\dag}_j\to\hat{S}^x_j+i\hat{S}^y_j$, $\hat{b}^{}_j\to\hat{S}^x_j-i\hat{S}^y_j$, and $\hat{n}^{b}_j\to\hat{S}^z_j+1/2$~\cite{TMatsubara_1956,MACazalilla2011}.
For a bipartite lattice, after a further transformation $\hat{S}^{x(y)}_j\to(-1)^j\hat{S}^{x(y)}_j$, the Hamiltonian becomes 
\begin{equation}\label{eq:xxz}
    \hat{H}_{\rm XXZ} = J\sum_{\braket{i,j}} \left(
    \hat{S}^x_i\hat{S}^x_j + \hat{S}^y_i\hat{S}^y_j + \Delta_z\hat{S}^z_i\hat{S}^z_j\right) - h_z\sum_{j} \hat{S}^z_j, 
\end{equation}
where
\begin{equation}\label{xxzparams}
    J= 2t_{\rm eff},\quad
    \Delta_z = \frac{V_{\rm eff}}{2t_{\rm eff}},\quad
    h_z = -\epsilon_{\rm eff} -\frac{V_{\rm eff}}{2}z.
\end{equation}
Since the total number of itinerant electrons (holes) $\tilde{N}^c$ ($N_h=2L_x-\tilde{N}^c$) is conserved with the assumption that all holes form rung pairs, the particle number in the effective boson model is also conserved at $N_b=N_h/2=L_x-\tilde{N}^c/2$. 
As this particle number corresponds to the magnetization $M_{\rm XXZ}$ in the equivalent XXZ model, it is likewise conserved at $M_{\rm XXZ}=-L_x/2 + N_b = L_x/2-\tilde{N}^c/2$. 
Within a fixed magnetization sector, a uniform magnetic field ($h_z$) induces only a trivial energy shift and does not affect correlation functions. 
Note that Eq.~\eqref{eq:xxz} [Eq.~\eqref{eq:t-V_model}] is the Hamiltonian under periodic boundary conditions, where $h_z$ ($\epsilon_{\rm eff}$) is uniform; however, when using open boundary conditions in numerical calculations, it is necessary to pay attention to the difference in the coordination number $z$ at the edges. 
In the one-dimensional (1D) case, it is known that the XXZ model is exactly solvable by the Bethe ansatz, and the phase diagram as a function of $\Delta_z$ and $h_z$ is well established~\cite{CNYang_1996a,*CNYang_1996b,*CNYang_1996c,TGiamarchi2003book}. 

The single-particle correlation function $\langle \hat{b}^{\dag}_i \hat{b}^{}_j \rangle$ of the hardcore boson $t$-$V$ model corresponds to the transverse spin correlation $\langle \hat{S}_i^+ \hat{S}_j^- \rangle$ of the XXZ model. 
In the gapless phase of the 1D XXZ model, the correlation function exhibits a power-law decay, $\langle \hat{S}^+_{j+r}\hat{S}^-_j \rangle \sim (-1)^rr^{-1/(2K)}$, where $K$ denotes the Luttinger parameter~\cite{TGiamarchi2003book}. 
Since the behavior of the correlation function in the 1D XXZ model has been well studied, relating it to the pair correlation $\langle \hat{\Delta}_i^{\dag}\,\hat{\Delta}^{}_j \rangle$, where $\hat{\Delta}_j^{\dag} = \frac{1}{\sqrt{2}} (\hat{c}^{\dag}_{j,1,\uparrow}\hat{c}^{\dag}_{j,2,\downarrow}-\hat{c}^{\dag}_{j,1,\downarrow}\hat{c}^{\dag}_{j,2,\uparrow})$, is meaningful. 
For the ladder-type or bilayer $t$-$J$ model, the pair correlation function $\langle \hat{\Delta}_i^{\dag}\,\hat{\Delta}^{}_j \rangle$ is directly mapped onto the single-particle correlation $\langle \hat{b}^{\dag}_i \hat{b}^{}_j \rangle$ in the hardcore boson model. 
In the double Kondo lattice model, however, the correspondence between these correlations requires more careful treatment because the effective boson model is composed of both itinerant electrons and localized spins. 
Nevertheless, even in the double Kondo lattice model, the correlations $\langle \hat{b}^{\dag}_i \hat{b}^{}_j \rangle$ and $\langle \hat{\Delta}^{\dag}_i\,\hat{\Delta}^{}_j \rangle$ are essentially equivalent as follows. 
We define $\hat{P}$ as the projection operator onto the subspace $\mathcal{H}_n$ [see Eq.~\eqref{eq:subspace}]. 
Using the projection operator $\hat{P}$, we obtain the relations 
\begin{equation}
    \hat{P}\hat{\Delta}^{\dag}_j\hat{P} = \alpha\,\hat{b}^{}_j,\quad
    \hat{P}\hat{\Delta}^{}_j\hat{P} = \alpha^*\, \hat{b}^{\dag}_j,
\end{equation}
where
\begin{equation}
    \alpha = \langle g^{\rm (T)}|\hat{\Delta}^{\dag}_j|{\rm hh}\rangle.
\end{equation}
As long as the low-energy physics is well captured within $\mathcal{H}_n$, i.e., for $t_{\parallel} \ll \Delta_{\rm s}, E_{\rm B}(<J_{\perp})$, the pair correlation function is proportional to the boson single-particle correlation function:
\begin{equation}\label{eq:correspondence_pair_corr}
    \langle \hat{\Delta}_i^{\dag}\,\hat{\Delta}^{}_j \rangle 
    \simeq |\alpha|^2 \langle \hat{b}^{}_i \hat{b}^{\dag}_j \rangle = |\alpha|^2 \langle \hat{b}^{\dag}_j \hat{b}^{}_i \rangle.
\end{equation}
Therefore, from this relation, the pair correlation function in the double Kondo lattice model can be characterized by the equivalent spin correlation function in the XXZ model and the renormalization factor $\alpha$. 
In particular, for $t_{\perp}=0$, one finds $\alpha = \cos\frac{\theta}{2}$.

\begin{figure}[t]
\begin{center}
\includegraphics[width=\columnwidth]{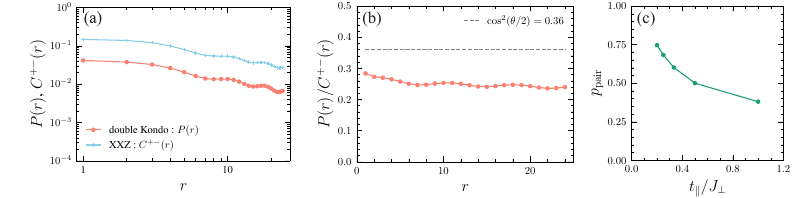} 
\caption{(a) Pair correlation function $P(r)=\langle \hat{\Delta}_{j+r}^{\dag}\,\hat{\Delta}_j \rangle$ (red) of the 1D double Kondo lattice model with $t_{\perp}=0$ at $n_h=N_h/(2L_x)=0.125$ and transverse spin correlation function $C^{+-}(r)=(-1)^r\langle \hat{S}^+_{j+r}\hat{S}^-_{j} \rangle$ (light blue) of the corresponding XXZ chain at the magnetization $M_{\rm XXZ}/L_x = n_h-1/2 = -0.375$.
The parameters $(t_{\parallel}, t_{\perp},J_{\rm K})=(0.2,0,2.0)$ in units of $J_{\perp}$ and $U/t_{\parallel}=8$ are used, where $J$ and $\Delta_z$ for the XXZ chain are calculated through Eqs.~\eqref{eq:bosonparams} and \eqref{xxzparams}. 
(b) Ratio $P(r)/C^{+-}(r)$ along with the ideal value $|\alpha|^2=\cos^2(\theta/2)\approx0.36$ (dashed line) in the $t_{\parallel}/J_{\perp} \rightarrow 0$ limit [see Eq.~\eqref{eq:correspondence_pair_corr}]. 
(c) $t_{\parallel}/J_{\perp}$ dependence of the fraction of holes forming rung pairs $p_{\rm pair}$.}
\label{figsxxz}
\end{center}
\end{figure}

We numerically examine the relationship mentioned above using the density-matrix renormalization group (DMRG) method. 
In Fig.~\ref{figsxxz}(a), we compare the pair correlation function $P(r)=\langle \hat{\Delta}_{j+r}^{\dag}\,\hat{\Delta}_j \rangle$ of the 1D double Kondo lattice model with the transverse spin correlation function $C^{+-}(r)=(-1)^r\langle \hat{S}^+_{j+r}\hat{S}^-_{j} \rangle$ of the corresponding XXZ chain, where the same parameter set as in Fig.~2(c) of the main text is used. 
Even for $t_{\parallel}/J_{\perp}=0.2$, a clear proportional relationship is observed between $P(r)$ and $C^{+-}(r)$. 
If the mapping carried out in Sec.~\ref{sec:mapping} is exact, the constant of proportionality should ideally be $\cos^2(\theta/2)$.
However, as shown in Fig.~\ref{figsxxz}(b), the ratio $P(r)/C^{+-}(r)$ actually takes a slightly smaller value than the ideal one. 
This deviation can be understood from the fact that the fraction of holes forming rung pairs among all holes, $p_{\rm pair} := 2N_{\perp}^{\rm pair} / N_h$ with $N_{\perp}^{\rm pair} = \sum_j \big\langle \prod_{l}\prod_{\sigma}(1-\hat{n}^c_{j,l,\sigma}) \big\rangle$, exhibits $p_{\rm pair} \approx 0.75 < 1$ at $t_{\parallel}/J_{\perp}=0.2$ [see Fig.~\ref{figsxxz}(c)].
Even when $t_{\parallel}$ renders the ground state nontrivial, the clear proportional relationship between $P(r)$ and $C^{+-}(r)$ still strongly supports the validity of the effective picture based on the exact ground state of a single rung.

\section{Kondo lattice model bridged by more than two spins}

\begin{figure}[t]
\begin{center}
\includegraphics[width=\columnwidth]{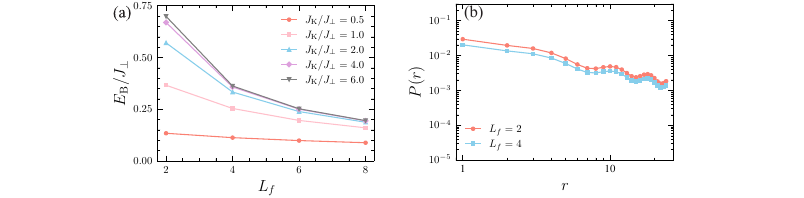}
\caption{(a) Binding energy $E_{\rm B}$ for a single rung with $L_f$ localized spins in the extended double Kondo lattice model. 
(b) Pair correlation functions $P(r)$ for $L_f=2$ and $L_f=4$ at $J_{\rm K}/J_\perp=2.0$, where $U/t_{\parallel}=8$ and $n_h=0.125$. 
We set $t_{\parallel}/J_{\perp}=1.0$ for $L_f=2$ and $t_{\parallel}/J_{\perp}=0.5$ for $L_f=4$.}
\label{fig:multi}
\end{center}
\end{figure}

The above discussion can be extended to other systems. 
As an example, we consider an extended double Kondo lattice model in which more than two localized spins bridge two itinerant-electron sites. 
When a unit cell contains $L_f$(=even integer) localized spins, the Hamiltonian along the rung direction is given by
\begin{align}\label{eq:extended model}
\hat{H}_{\perp} = \sum_{j} \Bigg[J_{\perp} \sum_{l=1}^{L_f-1}\hat{\bm{S}}^f_{j,l} \cdot \hat{\bm{S}}^f_{j,l+1}
+ J_{\rm K} \left(\hat{\bm{S}}^f_{j,1} \cdot \hat{\bm{S}}^c_{j,1}+\hat{\bm{S}}^f_{j,L_f} \cdot \hat{\bm{S}}^c_{j,2} \right) 
+ U\sum_{l=1}^2 \hat{n}^c_{j,l,\uparrow} \hat{n}^c_{j,l,\downarrow} 
\Bigg],
\end{align}
while that along the chain (in-plane) direction is given in Eq.~\eqref{eq:hopterm}. 
The case of $L_f=2$ corresponds to the normal double Kondo lattice model. 
The binding energies for a single rung ($L_x=1$) obtained by the DMRG method are shown in Fig.~\ref{fig:multi}(a). 
The binding energy $E_{\rm B}$ decreases as the bridge becomes longer, but remains nonzero.
We compare the pair correlation functions for $L_f=2$ and $L_f=4$ at $J_{\rm K}/J_\perp=2.0$ in Fig.~\ref{fig:multi}(b). 
We use $t_{\parallel}/J_{\perp}=1.0$ for $L_f=2$ and $t_{\parallel}/J_{\perp}=0.5$ for $L_f=4$, where the binding energies ($E_{\rm B}/t_{\parallel}$) of these two cases are comparable. 
We find that the pair correlation functions exhibit similar behavior for $L_f=2$ and $L_f=4$. 
Therefore, even when $L_f$ is different, the pair correlations similarly develop as long as the binding energies are comparable.
This demonstration suggests that our finding is not specific to the $L_f=2$ double Kondo lattice model. 

\section{Supplementary numerical results}

\subsection{Bond dimension and system-size dependence}

\begin{figure}[H]
\centering
\includegraphics[width=\columnwidth]{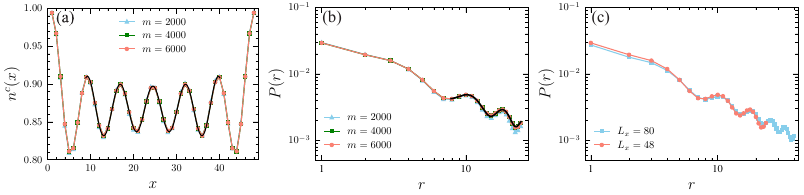}
\caption{Bond dimension ($m$) dependence of the (a) electron density $n^c(x)$ and (b) pair correlation function $P(r)$ for $L_x=48$, where the black solid lines are the fitting curves for $m=6000$. 
(c) Size ($L_x$) dependence of $P(r)$ for $m=6000$. 
Here, $n_h=0.125$, $J_{\rm K}/J_{\perp}=2$, $t_{\parallel}/J_{\perp}=1$, and $U/t_{\parallel}=8$ are used.}
\label{figs:bondsize}
\end{figure}

Here, we confirm the validity of our DMRG calculations. 
Figure~\ref{figs:bondsize}(a) shows the bond dimension $m$ dependence
of the electron density $n^c(x) = (1/2)\sum_{l,\sigma} \braket{ \hat{n}_{x, l,\sigma}}$. 
Similarly, Fig.~\ref{figs:bondsize}(b) shows the $m$ dependence of the pair correlation function $P(r)$. 
Both $n^c(x)$ and $P(r)$ converge sufficiently at $m\ge 4000$. 
In Fig.~\ref{figs:bondsize}(c), we compare $P(r)$ for $L_x=48$ with that for $L_x=80$. 
Since the tendencies of power-law decays are nearly identical, the characteristics of the pair correlation are sufficiently captured even at $L_x=48$, which is used in the main text. 

To obtain the decay exponent $K_{\rm C}$ of the charge-density-wave correlation, we fit $n^c(x)$ using~\cite{SRWhite2002} 
\begin{equation}
n^c(x) = A \frac{\cos(qx+\varphi)}{[\sin(\pi x/(L_x+1))]^\frac{K_{\rm C}}{2}}+n_0,
\end{equation}
where $A$, $\varphi$, and $q$ are the amplitude, phase shift, and wavenumber, respectively. 
Here, $n_0$ is treated as a fitting parameter. 
The data at $9 \le x \le 40$ are used in the fitting of $n^c(x)$. 
As in the main text, to obtain the decay exponent $K_{\rm SC}$, we fit the pair correlation function using $P(r) = A_0r^{-K_{\rm SC}}+A_1r^{-K_{\rm SC}^{\prime}}\cos(Qr +\phi)$~\cite{XLu2023,YShen2023PRB}, where the data at $8 \le r \le 24$ are used in the fitting. 
Here, the wavenumbers $Q$ for $P(r)$ and $q$ for $n^c(x)$ obtained from the fitting were nearly identical. 
The decay exponents obtained by the fitting curves for $m=6000$ shown in Figs.~\ref{figs:bondsize}(a) and \ref{figs:bondsize}(b) are $K_{\rm C} \approx 1.24$ and $K_{\rm SC} \approx 0.93$. 
Thus, one can find that the relationship $K_{\rm SC} \cdot K_{\rm C} \simeq 1$, which is a characteristic of the Luther-Emery liquid phase~\cite{MDolfi2015}, is approximately satisfied.

\FloatBarrier
\subsection{Spin correlation functions}

\begin{figure}[H]
\begin{center}
\includegraphics[width=\columnwidth]{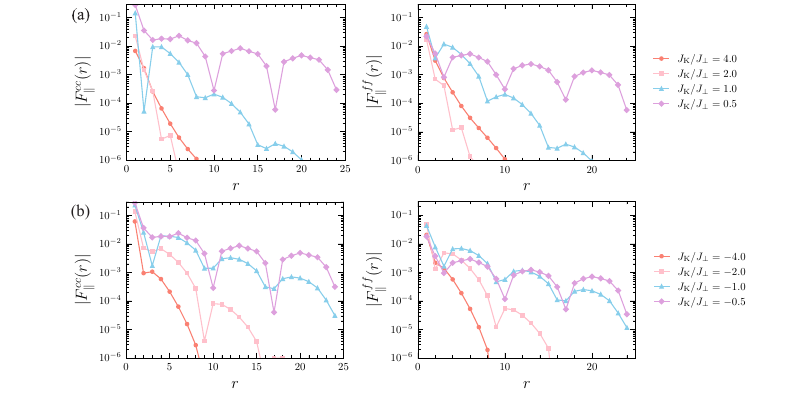}
\caption{Spin correlation functions along the chain-direction $F_{\parallel}^{cc}(r)$ (left panels) and $F_{\parallel}^{ff}(r)$ (right panels) for (a) $J_{\rm K}>0$ with $t_{\parallel}/J_{\perp}=1$ and (b) $J_{\rm K}<0$ with $t_{\parallel}/J_{\perp}=1/3$, where $n_h=0.125$ and $U/t_{\parallel}=8$.}
\label{figs:spinchain}
\end{center}
\end{figure}

In one dimension, a system is considered to be in a superconducting phase when a pair correlation function decays more slowly than other correlation functions. 
To confirm the dominance of the pair correlation in the double Kondo lattice model, we present the spin correlation functions along the chain direction
\begin{equation}
F_{\parallel}^{\mu\mu}(r) = \braket{\hat{\bm{S}}^{\mu}_{j+r,1} \cdot \hat{\bm{S}}^{\mu}_{j,1}} 
\end{equation}
in Fig.~\ref{figs:spinchain}, where we set $j=L_x/4$.
We find that the spin correlations exhibit clear exponential decays for $|J_{\rm K}|/J_\perp\gtrsim 2.0$, where the pair correlation functions are significantly enhanced with power-law decays. 
This demonstrates that the pair correlations are more dominant than the spin correlations at $|J_{\rm K}|/J_\perp\gtrsim 2.0$.
The exponential decay of the spin correlation function implies the opening of the spin gap. 
The different tendencies with respect to the sign of $J_{\rm K}$ originate from the differences in the spin gap [see Eq.~\eqref{eq:spin_gap}].

\subsection{Effect of the on-site interaction $U$}

\begin{figure}[H]
\begin{center}
\includegraphics[width=\columnwidth]{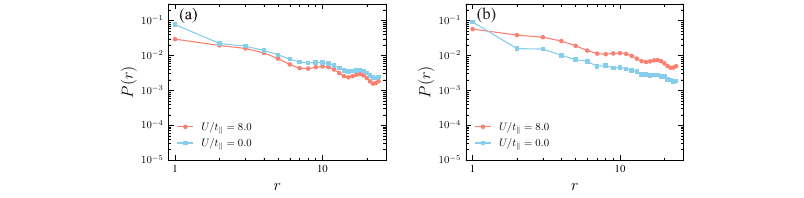}
\caption{Pair correlation functions with and without $U$ for (a) $J_{\rm K}/J_\perp=2.0$ with $t_{\parallel}/J_{\perp}=1.0$ and (b) $J_{\rm K}/J_\perp=-2.0$ with $t_{\parallel}/J_{\perp}=1/3$, where $n_h=0.125$.} 
\label{figs:U}
\end{center}
\end{figure}

Figure~\ref{figs:U} shows the pair correlation function with and without the on-site repulsion $U$. 
In contrast to the single-orbital Hubbard model, in which $U$ leads to the pairing glue~\cite{Scalapino2012}, the pair correlation functions in the double Kondo lattice model develop even in the absence of $U$. 
This behavior originates from the fact that the eigenstates of the four-site Kondo-Hubbard model without $t_{\perp}$ [Eqs.~\eqref{eq:E_200} and \eqref{eq:E_112M}] are independent of $U$, indicating that one obtains the binding energy as well as Eq.~\eqref{eq:binding_energy}. 
Hence, the nonvanishing binding energy supports the development of the pair correlation even for $U=0$. 

\bibliography{Reference}